\documentclass [useAMS]{mn2e}
\usepackage {graphicx}

\begin{document}

\title{Transfer of energy and angular momentum in magnetic coupling process }
\date{Accepted 0000 00 00. Received 0000 00 00}
\pubyear{0000} \volume{000}
\author[Ding-Xiong Wang, Wei-Hua Lei, Ren-Yi Ma]{Ding-Xiong
Wang$^*$, Wei-Hua Lei, Ren-Yi Ma \\
Department of Physics, Huazhong University of Science and
Technology, Wuhan,430074,China \\
$^*$Send offprint requests to: D.-X. Wang (dxwang@hust.edu.cn) }
\maketitle

\label{firstpage}

\begin{abstract}
The transfer of energy and angular momentum in the magnetic
coupling (MC) of a rotating black hole (BH) with its surrounding
accretion disc is discussed based on a mapping relation derived by
considering the conservation of magnetic flux with two basic
assumptions: (i)the magnetic field on the horizon is constant,
(ii) the magnetic field on the disc surface varies as a power law
with the radial coordinate of the disc. The following results are
obtained: (i) the transfer direction of energy and angular
momentum between the BH and the disc depends on the position of a
co-rotation radius relative to the MC region on the disc, which is
eventually determined by the BH spin; (ii) the evolution
characteristics of a rotating BH in the MC process without disc
accretion are depicted in a parameter space, and a series of
values of the BH spin are given to indicate the evolution
characteristics; (iii) the efficiency of converting accreted mass
into radiation energy of a BH-disc system is discussed by
considering coexistence of disc accretion and the MC process; (iv)
the MC effects on disc radiation and emissivity index are
discussed and it is concluded that they are consistent with the
recent \textit{XMM-Newton} observation of the nearby bright
Seyfert 1 galaxy MCG-6-30-15 with reference to a variety of
parameters of the BH-disc system.
\end{abstract}


\section{INTRODUCTION}

Recently, the magnetic coupling of a rotating black hole (BH) with
its surrounding disc has been investigated by certain authors
(Blandford 1999; Li 2000, Li 2002a, b, hereafter Li02a and Li02b,
respectively; Li {\&} Paczynski 2000; Wang, Xiao {\&} Lei 2002,
hereafter WXL). This coupling  can be regarded a variation of the
Blandford-Znajek (BZ) process, proposed two decades ago (Blandford
{\&} Znajek 1977). With closed magnetic field lines connecting a
rotating BH with the disc,  energy and angular momentum can be
transferred from the BH to the disc and henceforth this energy
mechanism is referred to as the magnetic coupling (MC) process.
The load in the MC process is the surrounding disc, which is much
better understood than the remote astrophysical load in the BZ
process. Energy and angular momentum are always transferred from
the BH to the unknown remote load in the BZ process, while the
transfer direction of energy and angular momentum in the MC
process depends not only on the angular velocity of the BH but
also on that of the disc where each closed field line penetrates.

Usually, the transfer of energy and angular momentum in the MC
process is stated as follows: 'If a BH rotates faster than its
surrounding disc, it exerts a torque on the disc and energy and
angular momentum are extracted from the BH and transferred to the
disc, and vice versa'. However, this statement is rather vague and
only applicable to the case that the closed magnetic field lines
are attached to the inner edge of the disc. In a more realistic
model the closed field lines are assumed to connect the BH horizon
with the disc by attaching a ring with inner and outer boundary as
shown in Fig.1, where $r_{in} $ and $r_{out} $ are the radii of
the inner and outer boundary of the MC region, respectively, and
$\theta _1 $ and $\theta _2 $ are the corresponding angular
coordinates on the BH horizon.


\begin{figure}
\begin{center}
\includegraphics[width=6cm]{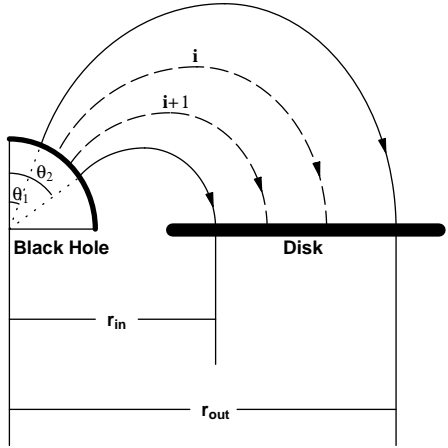}
\vspace{0.2cm} \caption{The poloidal magnetic field connecting a
rotating BH with its surrounding disc } \label{fig1}
\end{center}
\end{figure}

It is shown in WXL that the width of the MC region should not be
given at random, which can be determined by conservation of
magnetic flux in the context of general relativity. In this paper
the mapping relation derived in WXL between the angular coordinate
on the BH horizon and the radial coordinate on the disc is
modified by a boundary condition on the closed field lines
connecting the two loops in the equatorial plane of the BH. It
turns out that the transfer direction of energy and angular
momentum is determined by the position of the co-rotation radius
relative to the MC region, and ultimately depends on the BH spin
eventually. In addition, other MC effects, such as BH entropy
change, efficiency of releasing energy, disc radiation and
emissivity, also appear to be related intimately to the BH spin .
In particular, as argued in Li02b, the MC effects on emissivity
are consistent with the recent \textit{XMM-Newton} observation of
the nearby bright Seyfert 1 galaxy MCG-6-30-15. We find that a
variety of MC parameters are suitable for the above observation.

This paper is organized as follows. In Section 2 we give the basic
evolution equation of a BH in MC process, and derive a modified
mapping relation between the angular coordinate on the BH horizon
and the radial coordinate on the disc. In Section 3 the evolution
characteristics of a rotating BH in the MC process without disc
accretion are depicted in terms of the BH spin in a parameter
space, and a series of values of the BH spin are given to indicate
the evolution features. The sequence of these values are related
to the transfer of energy and angular momentum between the BH and
the disc, and guaranteed by the second law of BH thermodynamics
(Wald 1984). In Section 4 the efficiency of BH-disc systems in
converting accreted mass into radiation energy is discussed in
terms of the coexistence of disc accretion and the MC process. In
addition, we discuss MC effects on disc radiation and emissivity
index that are consistent with the recent \textit{XMM-Newton}
observation of the nearby bright Seyfert 1 galaxy MCG-6-30-15.
Finally in Section 5, we summarize our main results.

In order to facilitate the discussion of the correlation of BH
spin with MC effects in an analytic way, we make the following
assumptions:

(i) The disc is both stable and perfectly conducting, and the
closed magnetic field lines are frozen in the disc.

(ii) The disc is thin and Keplerian, and lies in the equatorial
plane of the BH with the inner boundary being at the marginally
stable orbit.

(iii) The magnetic field is assumed to be weak, and the effect of
instabilities of the disc and the magnetic field are ignored; The
magnetosphere is stationary, axisymmetric and force-free outside
the BH and the disc;.

(iv) The magnetic field is assumed to be constant on the horizon,
and to vary as a power law with the radial coordinate of the disc.

Throughout this paper the geometric units $G = c = 1$ are used.

\section{BASIC EQUATIONS FOR BH EVOLUTION AND MAPPING RELATION IN
THE MC PROCESS}

As argued by Macdonald {\&} Thorne (1982, hereafter MT), the
magnetic field on the horizon is brought and held by the
surrounding magnetized disc. So both disc accretion and the MC
process should be taken into account in our model for BH
evolution. Based on conservation of energy and angular momentum,
the basic equations for BH evolution with the coexistence of disc
accretion and the MC process are written as

\begin{equation}
\label{eq1} {dM} \mathord{\left/ {\vphantom {{dM} {dt}}} \right.
\kern-\nulldelimiterspace} {dt} = E_{ms} \dot {M}_D - P_{MC} ,
\end{equation}

\begin{equation}
\label{eq2} {dJ} \mathord{\left/ {\vphantom {{dJ} {dt}}} \right.
\kern-\nulldelimiterspace} {dt} = L_{ms} \dot {M}_D - T_{MC} .
\end{equation}

\noindent Combining equations (\ref{eq1}) and (\ref{eq2}) we have
the evolution of BH spin:

\begin{equation}
\begin{array}{l}
 da_*/dt = M^{ - 2}( {L_{ms} \dot {M}_D - T_{MC} }) \\ \\
 \quad\quad\quad\quad - 2M^{ - 1}a_ * ( {E_{ms} \dot {M}_D - P_{MC} }
 ).
\end{array}
 \label{eq3}
\end{equation}

Now we give a brief explanation for the quantities in equations
(\ref{eq1}) and (\ref{eq2}). $M$ and $J$ are mass and angular
momentum of the BH, respectively. $a_ * \equiv J \mathord{\left/
{\vphantom {J {M^2}}} \right. \kern-\nulldelimiterspace} {M^2}$ is
the dimensionless angular momentum of the BH, and referred to as
the BH spin. $E_{ms} $ and $L_{ms} $ are specific energy and
angular momentum corresponding to the inner edge of the thin disc,
i.e. the last stable circular orbit, respectively (Novikov {\&}
Thorne 1973). $\dot {M}_D $ is the accretion rate. At first sight,
$\dot {M}_D $ will be affected by the MC process owing to angular
momentum transferred between the rotating BH and the disc. We
derived the MC correction on $\dot {M}_D $ by considering the
conservation law of angular momentum and the angular momentum
transferred at the inner edge of the disc (see Appendix A in WXL).
Unfortunately, the modification to $\dot {M}_D $, equation
(\ref{eq41}) in WXL, is incorrect, since accretion rate cannot
depend on radius in a stationary flow, and the validity of the
following results of sections 5 and 6 in WXL is doubtful. In this
paper the MC effects on the accretion rate are not considered for
the following reasons.

(i) A constant accretion rate is required by the conservation of
mass in a stationary disc.

(ii) From equations (\ref{eq58}) and (\ref{eq59}) we find that the
viscous torque related to disc radiation can regulates itself
everywhere to counteract the effects of angular momentum
transferred into the disc on the accretion rate.

In the basic evolution equations (\ref{eq1}) and (\ref{eq2}),
$P_{MC} $ and $T_{MC} $ are the rates of extracting energy and
angular momentum from the rotating BH by the MC process, and
henceforth are referred to as the MC power and the MC torque,
respectively. We have derived the expressions for $P_{MC} $ and
$T_{MC} $ in WXL as follows:

\begin{equation}
\label{eq4} {P_{MC} } \mathord{\left/ {\vphantom {{P_{MC} } {P_0
}}} \right. \kern-\nulldelimiterspace} {P_0 } = 2a_ * ^2
\int_{\theta _1 }^{\pi \mathord{\left/ {\vphantom {\pi 2}} \right.
\kern-\nulldelimiterspace} 2} {\frac{\beta \left( {1 - \beta }
\right)\sin ^3\theta d\theta }{2 - \left( {1 - q} \right)\sin
^2\theta }},
\end{equation}

\begin{equation}
\label{eq5} {T_{MC} } \mathord{\left/ {\vphantom {{T_{MC} } {T_0
}}} \right. \kern-\nulldelimiterspace} {T_0 } = 4a_ * \left( {1 +
q} \right)\int_{\theta _1 }^{\pi \mathord{\left/ {\vphantom {\pi
2}} \right. \kern-\nulldelimiterspace} 2} {\frac{\left( {1 - \beta
} \right)\sin ^3\theta d\theta }{2 - \left( {1 - q} \right)\sin
^2\theta }},
\end{equation}

\noindent where $\theta $ is the angular coordinate on the horizon
varying from $\theta _1 $ to $\pi \mathord{\left/ {\vphantom {\pi
2}} \right. \kern-\nulldelimiterspace} 2$. $P_0 $ and $T_0 $ are
defined as

\begin{equation}
\label{eq6} \left\{ {\begin{array}{l}
 P_0 = \left\langle {B_H^2 } \right\rangle M^2 \approx B_4^2 M_8^2 \times
6.59\times 10^{44}erg \cdot s^{ - 1}, \\
 T_0 = \left\langle {B_H^2 } \right\rangle M^3 \approx B_4^2 M_8^3 \times
3.26\times 10^{47}g \cdot cm^2 \cdot s^{ - 2}, \\
 \end{array}} \right.
\end{equation}

\noindent and $B_4 $ and $M_8 $ are $\sqrt {\left\langle {B_H^2 }
\right\rangle } $ and $M$ in the units of $10^4gauss$ and $10^8M_
\odot $, respectively. The parameter $\beta \equiv {\Omega _D }
\mathord{\left/ {\vphantom {{\Omega _D } {\Omega _H }}} \right.
\kern-\nulldelimiterspace} {\Omega _H }$ is defined as the ratio
of the angular velocity on the thin disc to that on the horizon,
and we have the following expressions:

\begin{equation}
\label{eq7}
 \Omega_H=a_*/(2r_{_H}),\quad\quad r_{_H}=M(1+q), \quad\quad
 q=\sqrt{1-a_*^2},\\
\end{equation}

\noindent and

\begin{equation}
\label{eq8}
 \Omega_F=\Omega_D=\frac{1}{M(\chi^3+a_*)},\\
\end{equation}

\noindent where $r_{_H} $ is the radius of the horizon of a Kerr
BH. $\Omega _F $ is the angular velocity of the closed field line
connecting the BH and the disc, and $\Omega _F = \Omega _D $
arises from the freezing-in condition in the disc (MT).

In order to calculate $P_{MC} $ and $T_{MC} $ we should first
determine the mapping relation between the BH horizon and the
disc. Considering the flux tube consisting of two adjacent
magnetic surfaces ``$i$'' and ``$i + 1$'' as shown in Fig.1, we
have $\Delta \Psi _H = \Delta \Psi _D $ required by continuum of
magnetic flux, i.e.

\begin{equation}
\label{eq9} B_ \bot 2\pi \left( {\varpi \rho } \right)_{r = r_{_H}
} d\theta = - B_z 2\pi \left( {{\varpi \rho } \mathord{\left/
{\vphantom {{\varpi \rho } {\sqrt \Delta }}} \right.
\kern-\nulldelimiterspace} {\sqrt \Delta }} \right)_{\theta = \pi
\mathord{\left/ {\vphantom {\pi 2}} \right.
\kern-\nulldelimiterspace} 2} dr,
\end{equation}

\noindent where $B_ \bot $ and $B_z $ are the normal components of
magnetic field at the horizon and the disc, respectively, and

\begin{equation}
\label{eq10} \left( {\varpi \rho } \right)_{r = r_{_H} } = \left(
{\Sigma \sin \theta } \right)_{r = r_{_H} } = 2Mr_{_H} \sin \theta
,
\end{equation}

\begin{equation}
\label{eq11} \left( {{\varpi \rho } \mathord{\left/ {\vphantom
{{\varpi \rho } {\sqrt \Delta }}} \right.
\kern-\nulldelimiterspace} {\sqrt \Delta }} \right)_{\theta = \pi
\mathord{\left/ {\vphantom {\pi 2}} \right.
\kern-\nulldelimiterspace} 2} = \Sigma \mathord{\left/ {\vphantom
{\Sigma {\sqrt \Delta }}} \right. \kern-\nulldelimiterspace}
{\sqrt \Delta } = \alpha ^{ - 1}\rho ,
\end{equation}

\noindent where

\begin{equation}
\label{eq12} \alpha = \left( {{\rho \sqrt \Delta } \mathord{\left/
{\vphantom {{\rho \sqrt \Delta } \Sigma }} \right.
\kern-\nulldelimiterspace} \Sigma } \right)_{\theta = \pi
\mathord{\left/ {\vphantom {\pi 2}} \right.
\kern-\nulldelimiterspace} 2} = \sqrt {\frac{1 - 2\chi _{ms}^{ -
2} \xi ^{ - 1} + a_ * ^2 \chi _{ms}^{ - 4} \xi ^{ - 2}}{1 + a_ *
^2 \chi _{ms}^{ - 4} \xi ^{ - 2} + 2a_ * ^2 \chi _{ms}^{ - 6} \xi
^{ - 3}}} .
\end{equation}
In equation (\ref{eq12}) $\xi \equiv r \mathord{\left/ {\vphantom
{r {r_{ms} }}} \right. \kern-\nulldelimiterspace} {r_{ms} }$is a
dimensionless radial parameter defined in terms of the radius
$r_{ms} $ of the last stable circular orbit. Following Blandford
(1976) we assume that $B_z $ varies as

\begin{equation}
\label{eq13} B_z \propto \xi ^{ - n}.
\end{equation}

\begin{table*}
\begin{center}
\textbf{Table 1. \it{Some quantities related to $a_ * ^{eq} $}}

\begin{tabular}{ccccccc}
\hline \hspace{0.1cm}n& $a_ * ^{eq} $ & $\xi _{out} \left( {a_ *
^{eq} } \right)$ & $\xi _c \left( {a_ * ^{eq} } \right)$ & $R_\xi
$ & $R_P $ &
$R_T $  \\
\hline \hspace{0.1cm}1.1& 0.2924& 1.28273& 1.10957& 0.63274&
-1.26867&
-1.0104 \\
\hline \hspace{0.1cm}1.5& 0.2908& 1.29863& 1.11273& 0.60636&
-1.28244&
-1.0107 \\
\hline \hspace{0.1cm}3.0& 0.2835& 1.38780& 1.12758& 0.49025&
-1.35549&
-1.0123 \\
\hline
\end{tabular}
\end{center}
\label{tab1}
\end{table*}

\noindent Owing to the lack of knowledge of configuration of the
magnetic field in the gap region between the horizon and the inner
edge of the disc, we proposed the following assumptions in WXL:
(a) the inner boundary of the MC region is located at the inner
edge of the disc, with which the closed field lines connect the
horizon at $\theta = \pi \mathord{\left/ {\vphantom {\pi 2}}
\right. \kern-\nulldelimiterspace} 2$; (b) the strength of the
magnetic field at the horizon is equal to that at the inner edge
of the disc. However the former is probably greater than the
latter by numerical simulation in the disc (Ghosh {\&} Abramowicz
1997, and the references therein). Considering conservation of
magnetic flux in the inner boundary of the MC region, we propose
the following relation to replace assumption (b):

\begin{equation}
\label{eq14} 2\pi r_{_H} B_ \bot = 2\pi \varpi _D \left( {r_{ms} }
\right)B_z \left( {r_{ms} } \right),
\end{equation}

\noindent where $\varpi _D \left( {r_{ms} } \right)$ is the
cylindrical radius at the inner edge of a thin disc and reads

\begin{equation}
\label{eq15} \varpi _D \left( {r_{ms} } \right) = M\chi _{ms}^2
\sqrt {1 + \chi _{ms}^{ - 4} a_ * ^2 + 2\chi _{ms}^{ - 6} a_ * ^2
} .
\end{equation}
Equation (\ref{eq14}) implies conservation of magnetic flux
corresponding to the two loops of the same infinitesimal width,
and the ratio of $B_ \bot $ to $B_z \left( {r_{ms} } \right)$
varies from 1.8 to 3 for $0 < a_ * < 1$. Incorporating equations
(\ref{eq13}) and (\ref{eq14}) we have

\begin{equation}
\label{eq16} B_z = B_ \bot \left[ {{r_{_H} } \mathord{\left/
{\vphantom {{r_{_H} } {\varpi _D \left( {r_{in} } \right)}}}
\right. \kern-\nulldelimiterspace} {\varpi _D \left( {r_{in} }
\right)}} \right]\xi ^{ - n}.
\end{equation}
Combining equations (\ref{eq10})---(\ref{eq16}) into equation
(\ref{eq9}) we have

\begin{equation}
\label{eq17} \sin \theta d\theta = - \mbox{G}\left( {a_ * ;\xi ,n}
\right)d\xi ,
\end{equation}

\noindent where

\begin{equation}
\begin{array}{l}
 G( {a_ * ;\xi ,n} ) \\ \\
 =  \frac{\xi
^{1 - n}\chi _{ms}^2 \sqrt {1 + a_ * ^2 \chi _{ms}^{ - 4} \xi ^{ -
2} + 2a_ * ^2 \chi _{ms}^{ - 6} \xi ^{ - 3}} }{2\sqrt {\left( {1 +
a_ * ^2 \chi _{ms}^{ - 4} + 2a_ * ^2 \chi _{ms}^{ - 6} }
\right)\left( {1 - 2\chi _{ms}^{ - 2} \xi ^{ - 1} + a_ * ^2 \chi
_{ms}^{ - 4} \xi ^{ - 2}} \right)} }.
\end{array}
\label{eq18}
\end{equation}

\noindent Integrating equation (\ref{eq17}) and setting $\xi = \xi
_{in} = 1$ at $\theta = \pi \mathord{\left/ {\vphantom {\pi 2}}
\right. \kern-\nulldelimiterspace} 2$, we express the mapping
relation by

\begin{equation}
\label{eq19} \cos \theta = \int_1^\xi {\mbox{G}\left( {a_ * ;\xi
,n} \right)} d\xi ,
\end{equation}

\noindent and the outer boundary $\xi _{out} \equiv {r_{out} }
\mathord{\left/ {\vphantom {{r_{out} } {r_{ms} }}} \right.
\kern-\nulldelimiterspace} {r_{ms} }$ of the MC region can be
determined by the following equation:

\begin{equation}
\label{eq20} \cos \theta _1 = \int_1^{\xi _{out} } {\mbox{G}\left(
{a_ * ;\xi ,n} \right)} d\xi .
\end{equation}

\noindent From equation (\ref{eq20}) we find that $\xi _{out}
\left( {a_
* ;n,\theta _1 } \right)$ behaves as a non-monotonic function of
$a_ * $ for the given $n$ and $\theta _1 $ as shown in Fig.2. It
is noted that $\theta _1 $ is taken as $\pi \mathord{\left/
{\vphantom {\pi 6}} \right. \kern-\nulldelimiterspace} 6$
throughout this paper except where indicated otherwise.

\begin{figure}
\begin{center}
\centerline{\includegraphics[width=6cm]{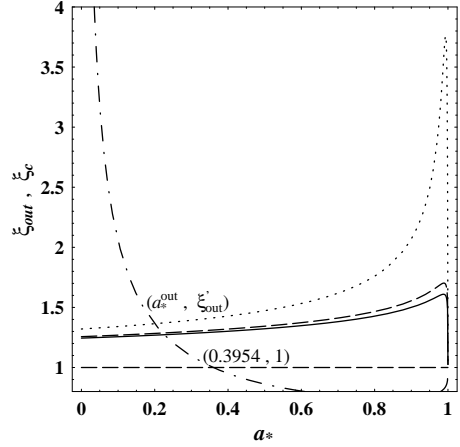}}
 \label{fig2}
 \caption{ The curves of $\xi _c $(dot-dashed line) and
$\xi _{out} $ versus $a_ * $ for $0 < a_ * < 1$ with $n =
1.1,\mbox{ 1.5}$ and $3.0$ in solid, dashed and dotted lines,
respectively.}
\end{center}
\end{figure}

The co-rotation radius $r_c $ is defined as the radius on the disc
where the angular velocity $\Omega _D $ of the disc is equal to
the BH angular velocity $\Omega _H $, i.e.

\begin{equation}
\label{eq21} {\Omega _D } \mathord{\left/ {\vphantom {{\Omega _D }
{\Omega _H }}} \right. \kern-\nulldelimiterspace} {\Omega _H } =
\frac{2\left( {1 + q} \right)}{a_
* }\left[ {\left( {\sqrt \xi \chi _{ms} } \right)^3 + a_ * } \right]^{ - 1}
= 1.
\end{equation}

From equation (\ref{eq21}) we can express $r_c $ in terms of a
parameter $\xi _c = {r_c } \mathord{\left/ {\vphantom {{r_c }
{r_{ms} }}} \right. \kern-\nulldelimiterspace} {r_{ms} }$ as
follows:

\begin{equation}
\label{eq22} \xi _c \left( {a_ * } \right) = {r_c }
\mathord{\left/ {\vphantom {{r_c } {r_{ms} }}} \right.
\kern-\nulldelimiterspace} {r_{ms} } = \chi _{ms}^{ - 2} \left( {1
- q} \right)^{ - 1 / 3}\left( {1 + q} \right).
\end{equation}

The parameter $\xi _c $ decreases monotonically with $a_ * $ is
shown by the dot-dashed line in Fig. 2, where there exist two
kinds of intersections: one is that the curve $\xi _c \left( {a_ *
} \right)$ intersects with each curve of $\xi _{out} \left( a_ *
;\theta _1 ,n \right)$, which is indicated by
  $(a_*^{out},\xi_{out}^{'})$; the other is that it intersects with
$\xi _{in} = 1$, which is indicated by $\left( {0.3594,\mbox{ }1}
\right)$. These intersections imply that the co-rotation radius
$r_c $ is located at the outer and inner boundaries of the MC
region for $a_ * = a_ * ^{out} $ and $a_ * ^{in} = 0.3594$,
respectively. The MC region is divided into two parts by $\xi _c
$: the inner MC region (henceforth IMCR) for $1 < \xi < \xi _c $
and the outer MC region (henceforth OMCR) for $\xi _c < \xi < \xi
_{out} $. Therefore energy and angular momentum are always
transferred by the closed magnetic field lines from the BH into
the  OMCR with $\Omega _D < \Omega _H $, while the transfer
direction reverses for IMCR with $\Omega _D > \Omega _H $. The
correlation of the BH spin with the transfer direction is given as
follows:

(i) For $0.3594 < a_ * < 1$ we have $\xi _c $ within the inner
edge, and the MC region is simply the OMCR with the transfer
direction from the  BH to the disc;

(ii) For $a_ * ^{out} < a_ * < 0.3594$ we have $1 < \xi _c < \xi
_{out} $, and the transfer direction is from the BH to the disc in
the OMCR, while it is from disc in the IMCR to the BH;

(iii) For $0 \le a_ * < a_ * ^{out} $ we have $\xi _c > \xi _{out}
$, and the MC region is simply the IMCR with the transfer
direction from the disc to the BH.

\section{BH EVOLUTION IN THE MC PROCESS WITHOUT DISC ACCRETION }
\subsection{Equilibrium spin of a BH in MC process without disc
accretion}

In order to highlight the MC effects on BH evolution we discuss a
specific case where disc accretion is absent. In this case a
rotating BH can attain its equilibrium spin $a_ * ^{eq} $ in the
MC process for $1 < \xi _c < \xi _{out} $. The basic evolution
equations (\ref{eq1}), (\ref{eq2}) and (\ref{eq3}) become

\begin{equation}
\label{eq23} {dM} \mathord{\left/ {\vphantom {{dM} {dt}}} \right.
\kern-\nulldelimiterspace} {dt} = - P_{MC} ,
\end{equation}

\begin{equation}
\label{eq24} {dJ} \mathord{\left/ {\vphantom {{dJ} {dt}}} \right.
\kern-\nulldelimiterspace} {dt} = - T_{MC} ,
\end{equation}

\begin{equation}
\label{eq25} {da_ * } \mathord{\left/ {\vphantom {{da_ * } {dt}}}
\right. \kern-\nulldelimiterspace} {dt} = - M^{ - 2}T_{MC} + 2M^{
- 1}a_ * P_{MC}.
\end{equation}

\noindent Setting ${da_ * } \mathord{\left/ {\vphantom {{da_ * }
{dt}}} \right. \kern-\nulldelimiterspace} {dt} = 0$, we have

\begin{equation}
\label{eq26} T_{MC} = 2Ma_ * P_{MC} .
\end{equation}
Substituting equations (\ref{eq4}), (\ref{eq5}) and (\ref{eq19})
into equation (\ref{eq26}), we can obtain $a_ * ^{eq} $ for the
given parameters $\theta _1 $ and $n$ by resolving the following
equation:

\begin{equation}
\label{eq27} \int_1^{\xi _{out} } {\frac{\left( {1 - \beta }
\right)\left( {1 - \beta + \beta q} \right)}{2\csc ^2\theta -
\left( {1 - q} \right)}} G\left( {a_ * ;\xi ,n} \right)d\xi = 0.
\end{equation}
In order to discuss the transportation of energy and angular
momentum at $a_
* ^{eq} $, we define the following ratios:

\begin{equation}
\label{eq28} R_P = {P_{MC}^{in} } \mathord{\left/ {\vphantom
{{P_{MC}^{in} } {P_{MC}^{out} }}} \right.
\kern-\nulldelimiterspace} {P_{MC}^{out} }, \quad R_T =
{T_{MC}^{in} } \mathord{\left/ {\vphantom {{T_{MC}^{in} }
{T_{MC}^{out} }}} \right. \kern-\nulldelimiterspace} {T_{MC}^{out}
},
\end{equation}

\noindent where $P_{MC}^{in} $ and $T_{MC}^{in} $ are the rates of
transferring energy and angular momentum in the IMCR,
respectively, and $P_{MC}^{out} $ and $T_{MC}^{out} $ are the
rates of transferring energy and angular momentum in OMCR,
respectively. By using the mapping relation (\ref{eq19}) these
quantities can be expressed as follows:

\begin{equation}
\label{eq29} P_{MC}^{in} = 2a_ * ^2 P_0 \int_1^{\xi _c }
{\frac{\beta \left( {1 - \beta } \right)G\left( {a_ * ;\xi ,n}
\right)}{2\csc ^2\theta - \left( {1 - q} \right)}} d\xi ,
\end{equation}

\begin{equation}
\label{eq30} P_{MC}^{out} = 2a_ * ^2 P_0 \int_{\xi _c }^{\xi
_{out} } {\frac{\beta \left( {1 - \beta } \right)G\left( {a_ *
;\xi ,n} \right)}{2\csc ^2\theta - \left( {1 - q} \right)}} d\xi ,
\end{equation}

\begin{equation}
\label{eq31} T_{MC}^{in} = 4a_ * T_0 \left( {1 + q}
\right)\int_1^{\xi _c } {\frac{\left( {1 - \beta } \right)G\left(
{a_ * ;\xi ,n} \right)}{2\csc ^2\theta - \left( {1 - q} \right)}}
d\xi ,
\end{equation}

\begin{equation}
\label{eq32} T_{MC}^{out} = 4a_ * T_0 \left( {1 + q}
\right)\int_{\xi _c }^{\xi _{out} } {\frac{\left( {1 - \beta }
\right)G\left( {a_ * ;\xi ,n} \right)}{2\csc ^2\theta - \left( {1
- q} \right)}} d\xi .
\end{equation}
From equations (\ref{eq19}), (\ref{eq22}),
(\ref{eq27})---(\ref{eq32}) we obtain some quantities related to
$a_ * ^{eq} $ as listed in Table 1.

From Table 1 we obtain the following results:

(i) Both $R_P $ and $R_T $ are less than $ - 1$, and these imply

\begin{equation}
\label{eq33}
\begin{array}{l}
P_{MC} \left( {a_ * ^{eq} } \right) = P_{MC}^{in} \left( {a_ *
^{eq} } \right) + P_{MC}^{out} \left( {a_ * ^{eq} } \right) < 0,\\
\\ T_{MC} \left( {a_ * ^{eq} } \right) = T_{MC}^{in} \left(
{a_
* ^{eq} } \right) + T_{MC}^{out} \left( {a_
* ^{eq} } \right) < 0,
\end{array}
\end{equation}

\noindent which means that the transportation of energy and
angular momentum in OMCR is dominated by that in IMCR, i.e. they
are transferred as a whole from the disc to the BH at equilibrium
spin $a_ * ^{eq} $.

(ii) Defining $R_\xi \equiv {\left( {\xi _c - 1} \right)}
\mathord{\left/ {\vphantom {{\left( {\xi _c - 1} \right)} {\left(
{\xi _{out} - \xi _c } \right)}}} \right.
\kern-\nulldelimiterspace} {\left( {\xi _{out} - \xi _c }
\right)}$ to indicate the ratio of the radial width of IMCR to
that of OMCR, we find that both $R_\xi $ and $a_ * ^{eq} $
decrease as the increasing $n$, while both $\xi _{out} \left( {a_
* ^{eq} } \right)$ and $\xi _c \left( {a_
* ^{eq} } \right)$ increase with increasing $n$. These results are related
to the conservation of magnetic flux with the fact that more
magnetic field is concentrated in IMCR for greater value of $n$.

\subsection{BH evolution in the MC process without disc accretion}

We can discuss the BH evolution in the corresponding parameter
space as proposed in WXL. Evolution equations (\ref{eq23}),
(\ref{eq24}) and (\ref{eq25}) can be rewritten as

\begin{equation}
\label{eq34}
\left\{ {\begin{array}{l}
 {dM} / {dt} = - P_{MC} = 2P_0 f\left( {a_ *
;n,\theta _1} \right), \\ \\
 f\left( {a_ * ;n,\theta _1 } \right) = - a_ * ^2 \int_1^{\xi _{out} }
{\frac{\beta \left( {1 - \beta } \right)G\left( {a_ * ;\xi ,n}
\right)}{2\csc ^2\theta - \left(
{1 - q}\right)}} d\xi , \\
 \end{array}} \right.
\end{equation}

\begin{equation}
\label{eq35} \left\{ {\begin{array}{l}
 {dJ} \mathord{\left/ {\vphantom {{dJ} {dt}}} \right.
\kern-\nulldelimiterspace} {dt} = - T_{MC} = 4T_0 h\left( {a_ *
;n,\theta _1 } \right), \\ \\
 h\left( {a_ * ;n,\theta _1 } \right) = - a_ * \left( {1 + q}
\right)\int_1^{\xi _{out} } {\frac{\left( {1 - \beta }
\right)G\left( {a_ * ;\xi ,n} \right)}{2\csc
^2\theta - \left( {1 - q} \right)}} d\xi , \\
 \end{array}} \right.
\end{equation}

\begin{equation}
\label{eq36} \left\{ {\begin{array}{l}
 {da_ * } \mathord{\left/ {\vphantom {{da_ * } {dt}}} \right.
\kern-\nulldelimiterspace} {dt} = 4T_0 M^{ - 2}g\left( {a_ *
;n,\theta _1 } \right), \\ \\
 g\left( {a_ * ;n,\theta _1 } \right) = - a_ * \left( {1 + q}
\right)\times\\
\quad\quad\quad\quad\quad\quad\quad\quad \int_1^{\xi _{out} }
{\frac{\left( {1 - \beta } \right)\left( {1 - \beta + \beta q}
\right)G\left( {a_ * ;\xi ,n} \right)}{2\csc ^2\theta -
\left( {1 - q} \right)}} d\xi . \\
 \end{array}} \right.
\end{equation}

From equations (\ref{eq34})---(\ref{eq36}), we find that $f\left(
{a_ * ;n,\theta _1 } \right)$, $h\left( {a_ * ;n,\theta _1 }
\right)$ and $g\left( {a_ * ;n,\theta _1 } \right)$ have the same
signs as those of ${dM} \mathord{\left/ {\vphantom {{dM} {dt}}}
\right. \kern-\nulldelimiterspace} {dt}$, ${dJ} \mathord{\left/
{\vphantom {{dJ} {dt}}} \right. \kern-\nulldelimiterspace} {dt}$
and ${da_ * } \mathord{\left/ {\vphantom {{da_ * } {dt}}} \right.
\kern-\nulldelimiterspace} {dt}$, respectively. Setting $f\left(
{a_ * ;n,\theta _1 } \right) = 0$, $h\left( {a_ * ;n,\theta _1 }
\right) = 0$ and $g\left( {a_ * ;n,\theta _1 } \right) = 0$ for
$\theta _1 = \pi \mathord{\left/ {\vphantom {\pi 6}} \right.
\kern-\nulldelimiterspace} 6$, we have three characteristic curves
in the two-dimension space consisting of the parameters $a_ * $
and $n$ as shown in Fig. 3, where each black dot with an arrowhead
represents one BH evolution state. From left to right, the
parameter space is divided into four regions by these three
curves, and the signs of the rates of change of $M$, $J$ and $a_ *
$ in these four regions are listed in Table 2.

\begin{center}
\begin{figure}
\centerline{\includegraphics[width=6cm]{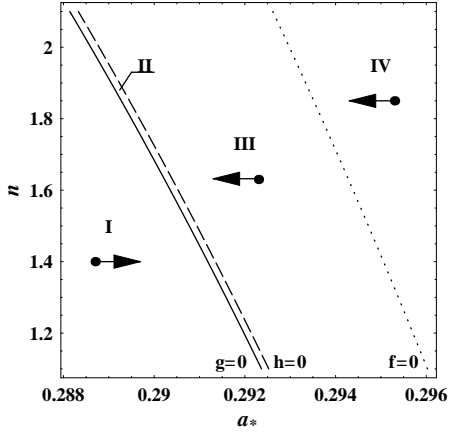}} \label{fig3}
\caption{Parameter space for BH evolution in the MC process with
}$1.1 \le n \le 2.1$ and $\mbox{0.}2881 < a_
* < 0.2960$\textbf{.}
\end{figure}
\end{center}

\begin{table}
\centerline{\textbf{Table 2.} {\em The signs of $dM/dt$, $dJ/dt$,
and $da_*/dt$ }} \centerline{\hspace{0.6cm} \em in the regions of
parameter space}
\begin{center}
\vspace{0cm}
\begin{tabular}{ccccc}
\hline \\
Region & \hspace{-0.4cm}Position   & \hspace{-0.4cm}$da_*/dt$   &   $dJ/dt$    &   $dM/dt$  \\
\hline \\
I    &   \hspace{-0.4cm} left of $g=0$        &  \hspace{-0.4cm}$>0$   &   $>0$  &   $>0$  \\
II   & \hspace{-0.4cm}between $g=0$ and $h=0$ &  \hspace{-0.4cm}$<0$   &   $>0$  &   $>0$  \\
III  & \hspace{-0.4cm}between $h=0$ and $f=0$ &  \hspace{-0.4cm}$<0$   &   $<0$  &   $>0$  \\
IV   & \hspace{-0.4cm}right of $f=0$          &  \hspace{-0.4cm}$<0$   &   $<0$  &   $<0$  \\
\hline
\end{tabular}
\end{center}
\end{table}

From Fig.3 we find that $P_{MC} $ changes its sign from negative
to positive at $a_ * ^P $ ($f = 0)$ and $T_{MC} $ does it at $a_ *
^T $ ($h = 0)$. The inequality, $a_ * ^T < a_ * ^P $, holds for
all possible values of $n$, because the curve $h = 0$ is located
on the left of $f = 0$ as shown in Fig.3\textbf{.} This result
means that $P_{MC} $ is negative with the positive $T_{MC} $ in
the value range $a_ * ^T < a_ * < a_ * ^P $. However, it does not
mean that the transfer direction of energy is opposite to that of
angular momentum in the IMCR or in the OMCR. In fact, as argued
above, the transfer direction of energy is always the same as that
of angular momentum either in the IMCR or in the OMCR, and the
opposite signs of $P_{MC} $ and $T_{MC} $ rest in the fact that
$P_{MC}^{out} $ and $T_{MC}^{in} $ are dominated by $P_{MC}^{in} $
and $T_{MC}^{out} $ for the above value range of the BH spin,
respectively. We can explain this order of the BH spin by using
the second law of BH thermodynamics in the next section. The order
of the above specific values of BH spin is given as follows:

\begin{equation}
\label{eq37} 0 < a_ * ^{out} < a_ * ^{eq} < a_ * ^T < a_ * ^P < a_
* ^{in} < 1.
\end{equation}

We can use the parameter space to determine the evolution
characteristics of a rotating BH in the MC process, provided that
the initial BH spin $a_ * $ and the power law index $n$ are given.
For example a BH with initial spin $a_ * ^P < a_ * < 1$ will
evolve to $a_ * ^{eq} $ and reach the curve $g = 0$ eventually by
passing region IV, III and II one after another. So we can
determine its evolution characteristics in each evolution stage
represented by the corresponding region in Fig.3.

The status of MC region and the signs of $P_{MC} $ and $T_{MC} $
for different value range of $a_ * $ are shown in Table 3.

\subsection{ BH entropy change in the MC process without disc
accretion}

Entropy $S_H $ of a Kerr BH can be expressed as (Wald 1984;
Thorne, Price {\&} Macdonald 1986)

\begin{equation}
S_H = 2\pi M^2(1 + q).
 \label{eq380}
 \end{equation}

\noindent From equations (\ref{eq23}) and (\ref{eq24}) we derive
the rate of change of $S_H $ as follows:

\begin{equation}
\label{eq38}
 \begin{array}{l}
 {dS_H }/{dt} = \Theta _H^{ - 1} \left( {\Omega _H
T_{MC} - P_{MC} } \right)\\ \\
 = 8\pi MP_0 (1 + q)^2\left( {q^{ -
1} - 1} \right)\int_{\theta _1 }^{\theta _2 } {\frac{\left( {1 -
\beta } \right)^2\sin ^3\theta d\theta }{2 - \left( {1 - q}
\right)\sin ^2\theta }},
\end{array}
\end{equation}

\noindent where $\Theta _H = \frac{q}{4\pi M(1 + q)}$ is the
temperature on the BH horizon. Inspecting equation (\ref{eq38}) we
find that ${dS_H } \mathord{\left/ {\vphantom {{dS_H } {dt}}}
\right. \kern-\nulldelimiterspace} {dt} > 0$ always holds. The
contribution to ${dS_H } \mathord{\left/ {\vphantom {{dS_H }
{dt}}} \right. \kern-\nulldelimiterspace} {dt}$ arises in two
parts: one part from $T_{MC} $, which is positive and negative for
$a_ * ^T < a_ * < 1$ and $0 < a_ * < a_ * ^T $, respectively, and
the other part from $ - P_{MC} $, which is positive and negative
for $0 < a_ * < a_ * ^P $ and $a_ * ^P < a_ * < 1$. It is noticed
that the two contributions are all positive for $a_ * ^T < a_ * <
a_ * ^P $. So the order $a_ * ^T < a_ * < a_ * ^P $ is guaranteed
by the second law of BH thermodynamics, otherwise the reverse
order will result in ${dS_H } \mathord{\left/ {\vphantom {{dS_H }
{dt}}} \right. \kern-\nulldelimiterspace} {dt} < 0$, and the
second law of BH thermodynamics will be violated. The
dimensionless rate of change of $S_H $ can be written as

\begin{equation}
\label{eq39}
 \begin{array}{l}
  \left( {rate} \right)_S \equiv \frac{{dS_H }
\mathord{\left/ {\vphantom {{dS_H } {dt}}} \right.
\kern-\nulldelimiterspace} {dt}}{\left( {{dS_H } \mathord{\left/
{\vphantom {{dS_H } {dt}}} \right. \kern-\nulldelimiterspace}
{dt}} \right)_0 } \\ \\
= (1 + q)^2\left( {q^{ - 1} - 1} \right)\int_1^{\xi _{out} }
{\frac{\left( {1 - \beta } \right)^2G\left( {a_ * ;\xi ,n}
\right)}{2\csc ^2\theta - \left( {1 - q} \right)}} d\xi,
 \end{array}
\end{equation}


\begin{table*}
\centerline{\textbf{Table 3.}{\it The signs of $P_{MC}$, $T_{MC}$
and the rates of change of BH parameters.}} \vspace{0cm}
\begin{center}
\begin{tabular}{ccccccccccc}
 \hline \\
 \hspace{0.3cm}$n$&
 \hspace{-1cm} \begin{tabular}{c} MC\\
  Region\end{tabular}&
 \hspace{-0.5cm}$a_*^{out}$&
 \hspace{-1cm}\begin{tabular}{c} MC\\ Region\end{tabular}&
 \hspace{-0.5cm}$a_*^{eq}$&
 \hspace{-1cm}\begin{tabular}{c} MC\\
  Region\end{tabular}&
 \hspace{-0.5cm}$a_*^T$&
 \hspace{-1cm}\begin{tabular}{c} MC\\
  Region \end{tabular} &
 \hspace{-0.5cm} $a_*^P$&
 \hspace{-1cm}\begin{tabular}{c} MC\\ Region\end{tabular}&
 \hspace{-0.5cm}$\geq 0.3594$ \\
 \hline \\
 \hspace{-0.2cm}\begin{tabular}{c}
  1.1\\1.5\\3.0\end{tabular}&
 \hspace{-1cm}\begin{tabular}{c} only\\OMCR\end{tabular}&
 \hspace{-1cm}\begin{tabular}{c} 0.2264\\0.2219\\0.1999\end{tabular} &
 \hspace{-1cm}\begin{tabular}{c} IMCR\\with\\OMCR\end{tabular} &
 \hspace{-1cm}\begin{tabular}{c} 0.2924\\0.2908\\0.2835\end{tabular}  &
 \hspace{-1cm}\begin{tabular}{c} IMCR \\ with \\ OMCR \end{tabular} &
 \hspace{-1cm} \begin{tabular}{c} 0.2925 \\ 0.2909 \\ 0.2837
   \end{tabular} &
 \hspace{-1cm}\begin{tabular}{c} IMCR\\with\\OMCR\end{tabular}&
 \hspace{-1cm}\begin{tabular}{c} 0.2960\\0.2947\\0.2890\end{tabular} &
 \hspace{-1cm}\begin{tabular}{c} IMCR \\ with \\ OMCR \end{tabular}  &
 \hspace{-1cm}\begin{tabular}{c} only \\ IMCR \end{tabular} \\
 \hline \\
 \hspace{0.3cm}$P_{MC}$&\hspace{-0.5cm}$<0$&\hspace{-0.5cm}$<0$&\hspace{-0.5cm}$<0$&\hspace{-0.5cm}$<0$&\hspace{-0.5cm}$<0$&\hspace{-0.5cm}$<0$&\hspace{-0.5cm}$<0$&\hspace{-0.5cm}$=0$&\hspace{-0.5cm}$>0$ &\hspace{-0.5cm} $>0$ \\
 \hspace{0.3cm}$T_{MC}$ &\hspace{-0.5cm} $<0$ &\hspace{-0.5cm} $<0$ &\hspace{-0.5cm}  $<0$ & \hspace{-0.5cm}$<0$  & \hspace{-0.5cm}$<0$ & \hspace{-0.5cm}$=0$ & \hspace{-0.5cm}$>0$ & \hspace{-0.5cm}$>0$
     & \hspace{-0.5cm}$>0$ & \hspace{-0.5cm}$>0$ \\
 \hspace{0.3cm}$dM/dt$ & \hspace{-0.5cm}$>0$ & \hspace{-0.5cm}$>0$ & \hspace{-0.5cm} $>0$ & \hspace{-0.5cm}$>0$  &\hspace{-0.5cm} $>0$ &\hspace{-0.5cm} $>0$ & \hspace{-0.5cm}$>0$ & \hspace{-0.5cm}$=0$
     & \hspace{-0.5cm}$<0$ & \hspace{-0.5cm}$<0$ \\
 \hspace{0.3cm}$dJ/dt$ & \hspace{-0.5cm}$>0$ &\hspace{-0.5cm} $>0$ & \hspace{-0.5cm} $>0$ &\hspace{-0.5cm} $>0$  & \hspace{-0.5cm}$>0$ & \hspace{-0.5cm}$=0$ & \hspace{-0.5cm}$<0$ &\hspace{-0.5cm} $<0$
     &\hspace{-0.5cm} $<0$ & \hspace{-0.5cm}$<0$ \\
 \hspace{0.3cm}$da_*/dt$ & \hspace{-0.5cm}$>0$ & \hspace{-0.5cm}$>0$ & \hspace{-0.5cm} $>0$ &\hspace{-0.5cm} $=0$  & \hspace{-0.5cm}$<0$ & \hspace{-0.5cm}$<0$ & \hspace{-0.5cm}$<0$ &
      \hspace{-0.5cm} $<0$ & \hspace{-0.5cm}$<0$ & \hspace{-0.5cm}$<0$ \\
\hline
\end{tabular}
\end{center}
\end{table*}

\noindent where $\left( {{dS_H } \mathord{\left/ {\vphantom {{dS_H
} {dt}}} \right. \kern-\nulldelimiterspace} {dt}} \right)_0 \equiv
8\pi MP_0 \approx 1.07\times 10^{60}B_4^2 M_8^3erg \cdot K^{ - 1}
\cdot s^{ - 1}$. From equation (\ref{eq39}) we obtain the curves
of $\left( {rate} \right)_S $\textbf{ }versus $a_ * $ for
different values of $n$ as shown in Fig. 4.


\begin{figure}
\begin{center}
{\includegraphics[width=5.5cm]{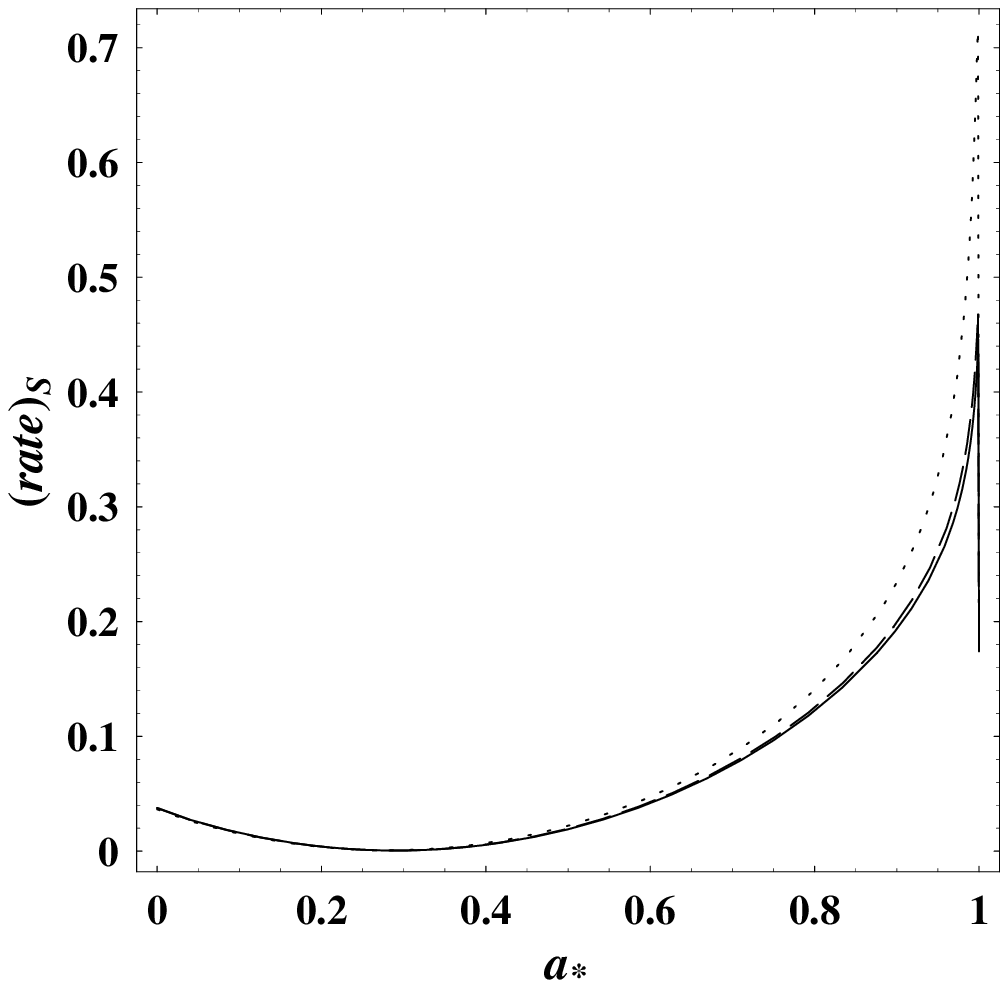}
 \includegraphics[width=5.5cm]{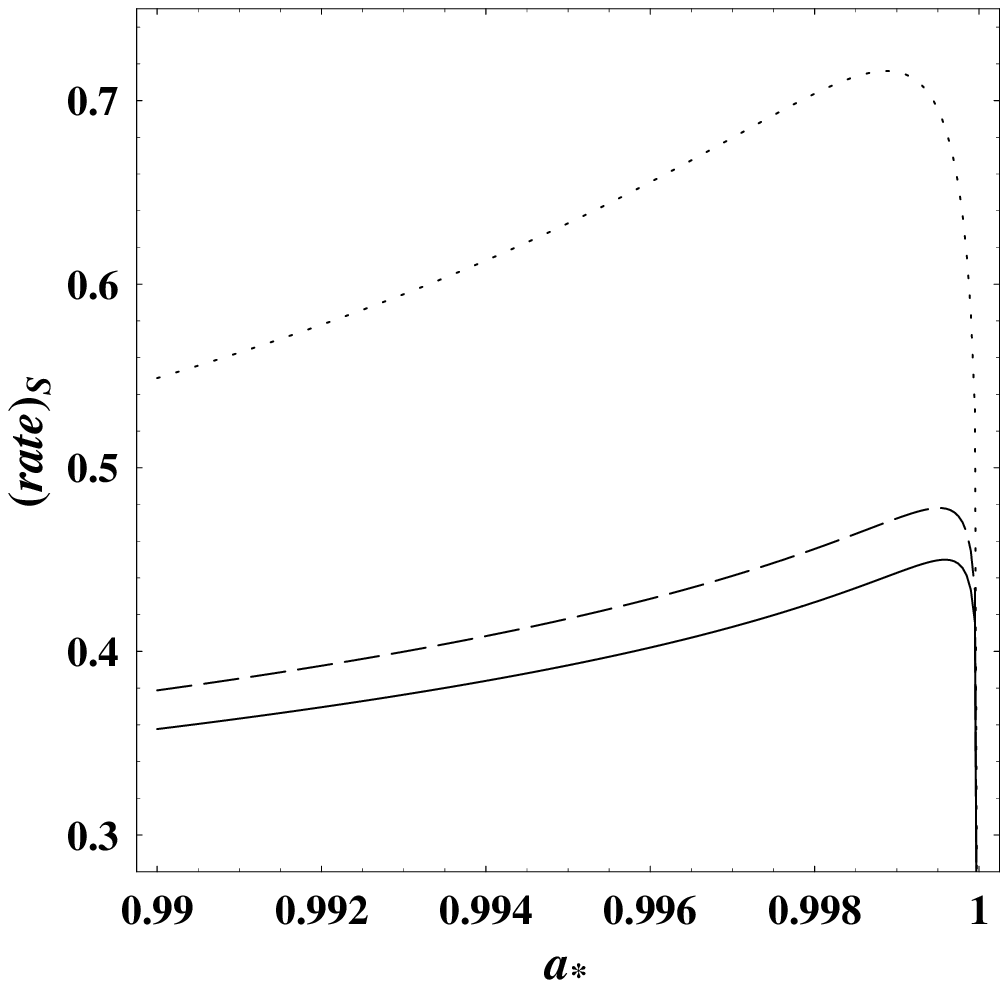}}
 \caption{ The curves of $\left( {rate} \right)_S $ versus $a_ *
$ with $n = 1.1,\mbox{ 1.5}$ and $3.0$ in solid, dashed and dotted
lines, respectively, with the upper panel for $0 < a_ * < 1$, and
the lower panel for $0.99 < a_
* < 1$.}
\label{fig4}
\end{center}
\end{figure}

From Fig.4 we find that $\left( {rate} \right)_S $ varies
non-monotonically as $a_ * $, attaining its minimum $\left( {rate}
\right)_S^{\min } $ at $a_
* ^{\min } $ and maximum $\left( {rate} \right)_S^{\max } $ at $a_ * ^{\max
} $, respectively. These limit values corresponding to the
different values of $n$ are listed in Table 4.

\begin{table}
\centerline{\textbf{Table 4.}{\em The values of
$(rate)_{_S}^{min}$, $(rate)_{_S}^{max}$}}
\centerline{\hspace{0.8cm} \em and the corresponding BH spin}
\begin{center}
\begin{tabular}{ccccc}
\hline \\
 $n$  &  $(a_*^{^S})_{min}$  &  $(rate)_{_S}^{min}$  &  $
 (a_*^{^S})_{max}$   &  $(rate)_{_S}^{max}$ \\
 \hline \\
 1.1  &  0.2909  &  0.0005382   &  0.99959  &  0.44995  \\
 1.5  &  0.2892  &  0.0005794   &  0.99960  &  0.47772  \\
 3.0  &  0.2812  &  0.0008010   &  0.99887  &  0.71611  \\
 \hline
\end{tabular}
\end{center}
\end{table}

From Table 4 we find that $\left( {a_ * ^S } \right)_{\min } $ is
located between $a_ * ^{out} $ and $a_ * ^{eq} $, which
corresponds to increasing $M$, $J$ and $a_ * $ with $P_{MC} < 0$
and $T_{MC} < 0$, while $\left( {a_ * ^S } \right)_{\max } $ is
very close to unity, corresponding to the decreasing $M$, $J$ and
$a_ * $ with $P_{MC} > 0$ and $T_{MC} > 0$. From Fig.4 we notice
that $\left( {rate} \right)_S $ decreases rapidly when $a_ * >
\left( {a_ * ^S } \right)_{\max } $, and we shall explain this
result in Section 5.

\section{MC EFFECTS ON THE BLACK HOLE-DISC SYSTEM}

Now we are going to discuss some evolution characteristics in MC
process with disc accretion (henceforth MCDA). The interaction
between a rotating BH with a disc is crucial for at least two
reasons:

(i) The magnetic field on the BH horizon is brought out and
maintained by the surrounding magnetized disc;

(ii) The transfer of energy and angular momentum between the BH
and the disc will remarkably affect not only BH evolution but also
  disc radiation.

\begin{figure*}
\begin{center}
{\includegraphics[width=4.1cm]{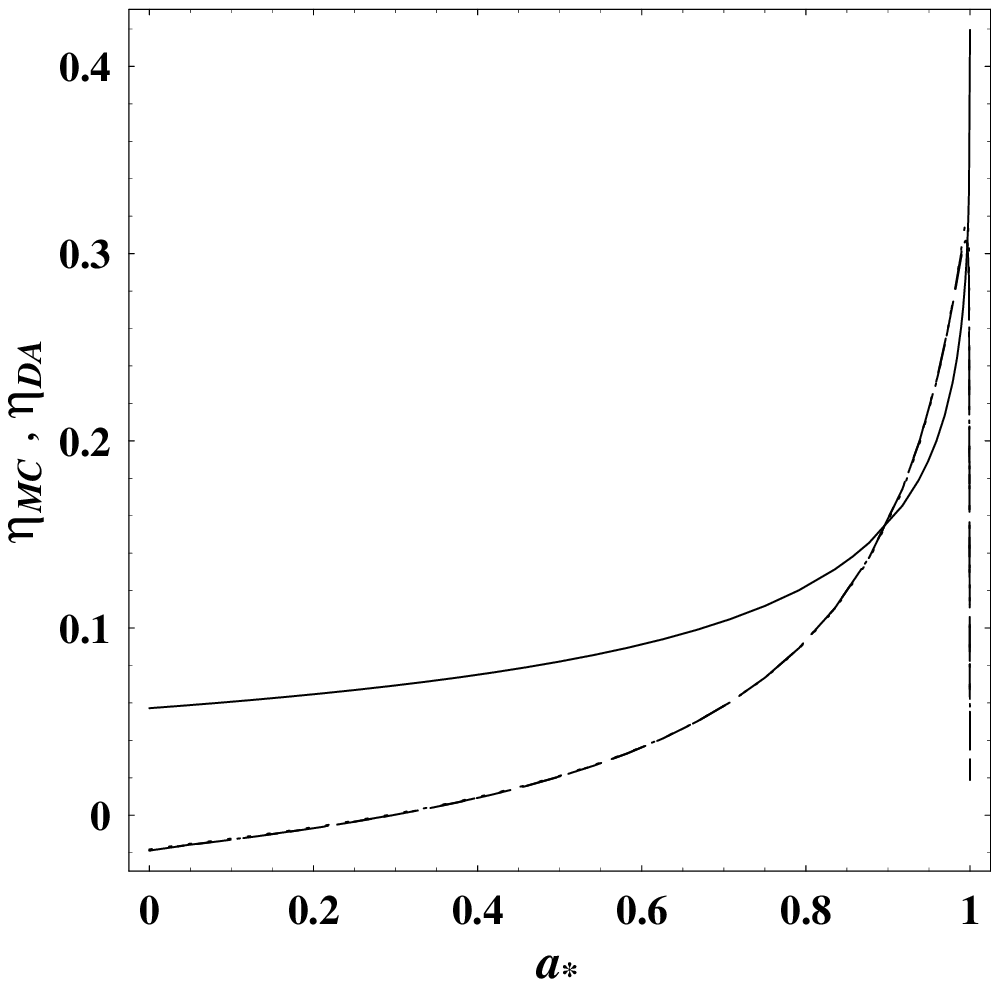}
 \hfill
\includegraphics[width=4.5cm]{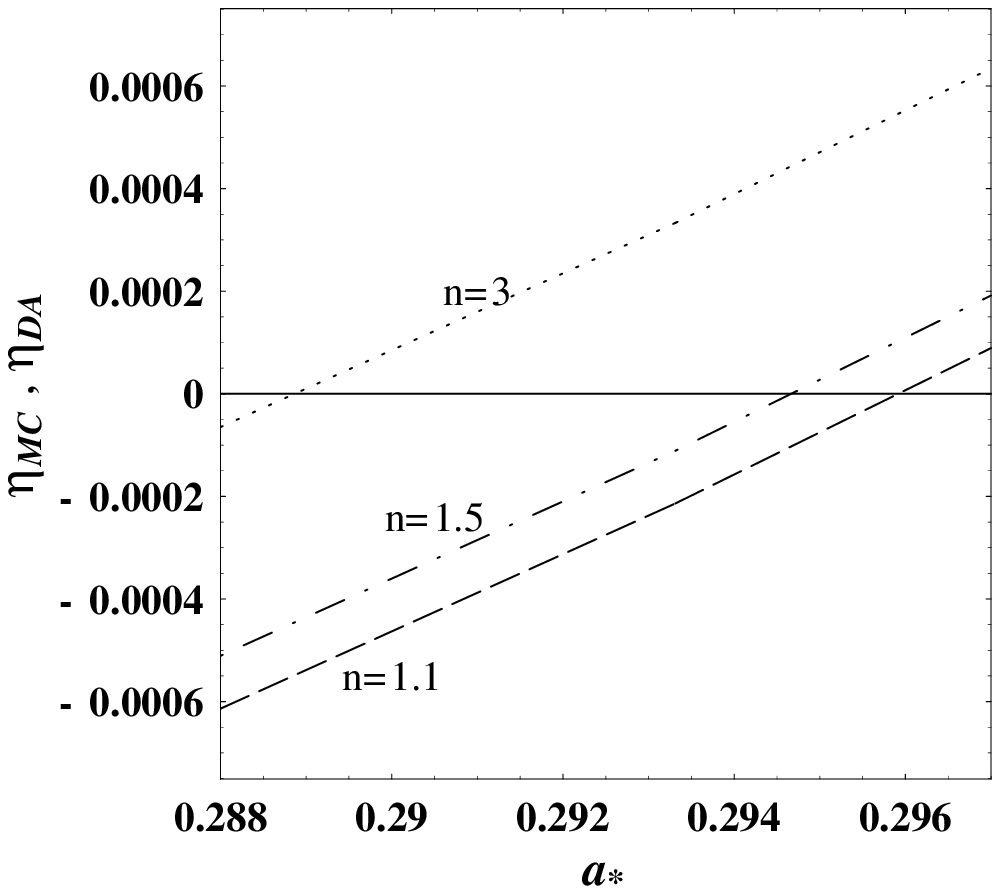}
\hfill
\includegraphics[width=4.5cm]{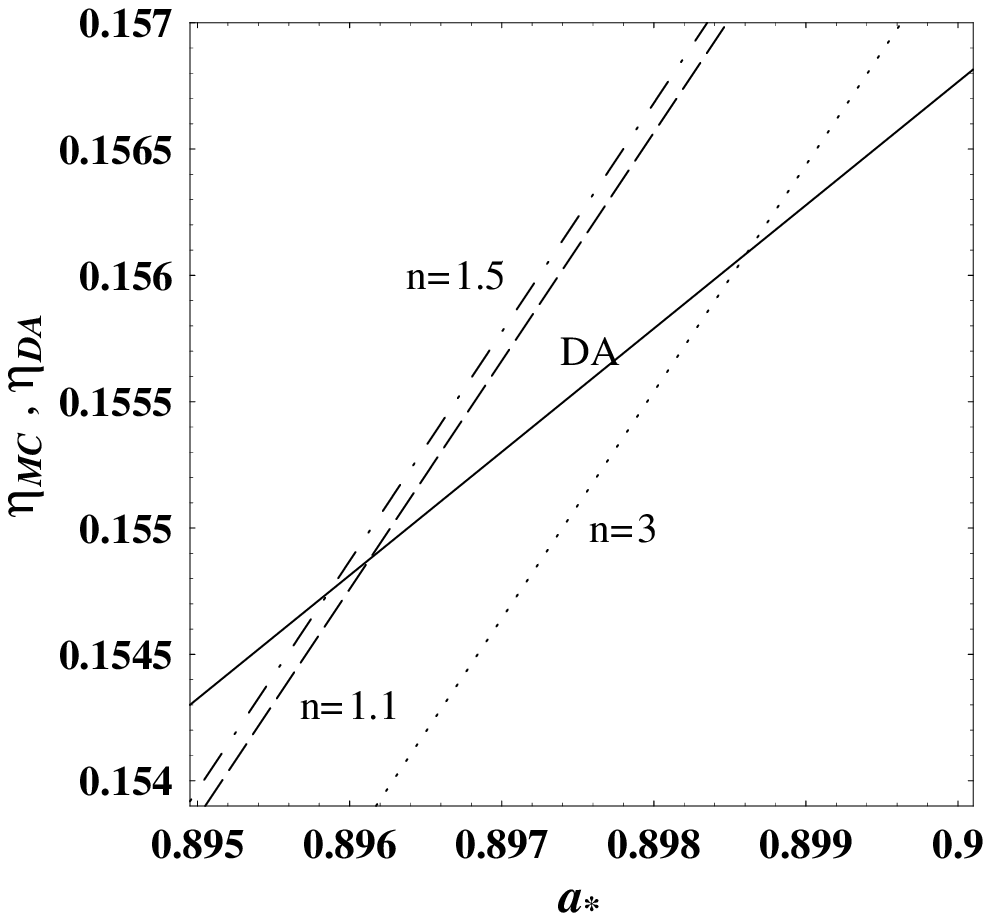}
 \hfill
 \includegraphics[width=4.25cm]{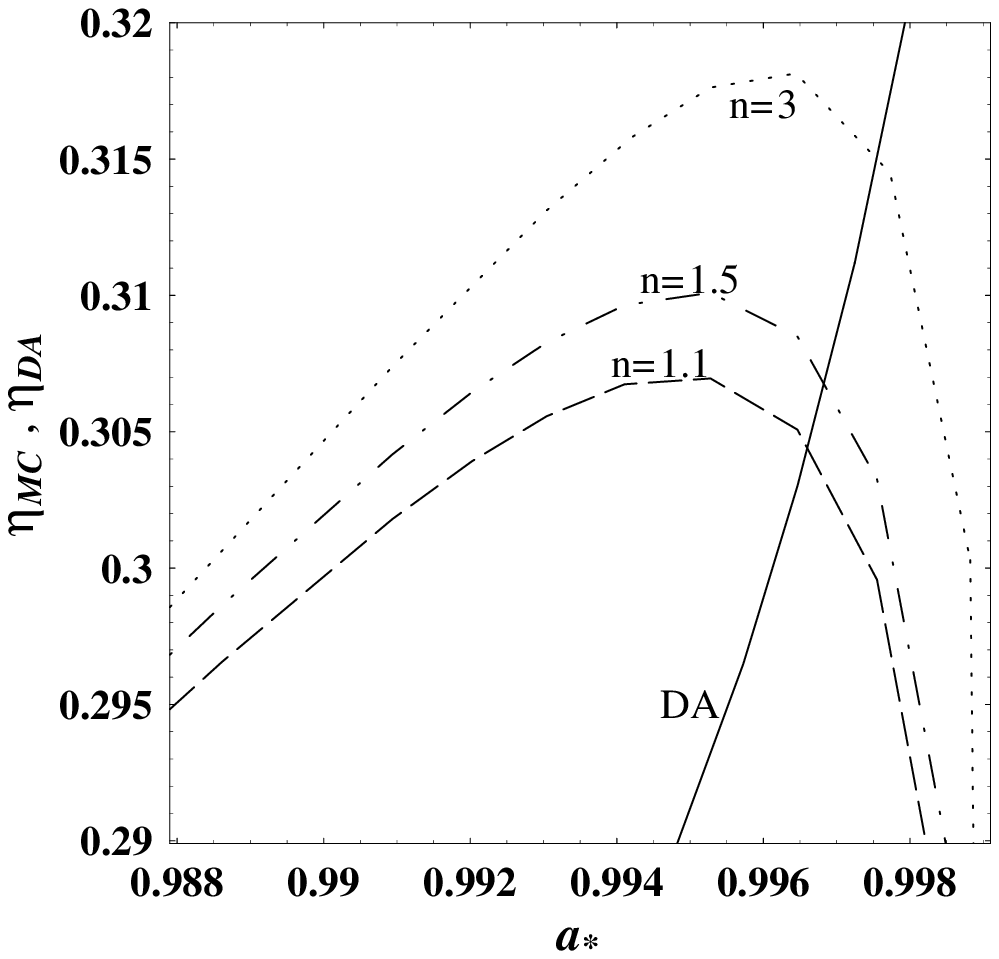}\\
 \centerline{\hspace{0.8cm}(a)\hspace{4.3cm}(b)\hspace{4.2cm}(c)\hspace{4.1cm}(d)}}
 \caption{The curves of $\eta _{_{DA}} $
(solid line) and $\eta _{_{MC}} $ versus $a_ * $ with $n = 1.1$,
$1.5$ and $3.0$ in dashed, dot-dashed and dotted lines,
respectively.(a) $0 < a_* < 1$, (b) $0.288 < a_* < 0.297$, (c)
$0.895 < a_* < 0.900$, (d) $0.988 < a_* < 0.999$}
 \label{fig5}
 \end{center}
 \end{figure*}


\begin{table*}
\centerline{\textbf{Table 5}--{\em Some specific value ranges of
$\eta_{_{MC}}$ and corresponding values of $a_*$}}
\begin{center}
\begin{tabular}{cccc}
 \hline
 $n$ &
 \begin{tabular}{ccc}  & Value range of $a_*$ & \\ \hline
 $\eta_{_{MC}}<0$\hspace{0.9cm} &  $0<\eta_{_{MC}}<\eta_{_{DA}}$ &
 \hspace{0.4cm} $\eta_{_{MC}}>\eta_{_{DA}}$ \end{tabular} &
 $(\eta_{_{MC}})_{max}$  & $(a_*^\eta)_{max}$ \\
 \hline \\
1.1 & \begin{tabular}{ccc} (0,0.2960) & (0.2960,0.8977),
(0.9966,1) & (0.8977,0.9966) \end{tabular} & 0.3070 & 0.9949 \\
1.5 & \begin{tabular}{ccc} (0,0.2947) & (0.2947,0.8974),
(0.9969,1) & (0.8974,0.9969) \end{tabular} & 0.3101 & 0.9951 \\
3.0 & \begin{tabular}{ccc} (0,0.2890) & (0.2890,0.9002),
(0.9976,1) & (0.9002,0.9976) \end{tabular} & 0.3182 & 0.9962 \\
 \hline
\end{tabular}
\end{center}
\end{table*}


\begin{figure*}
\begin{center}
 {\includegraphics[width=5.2cm]{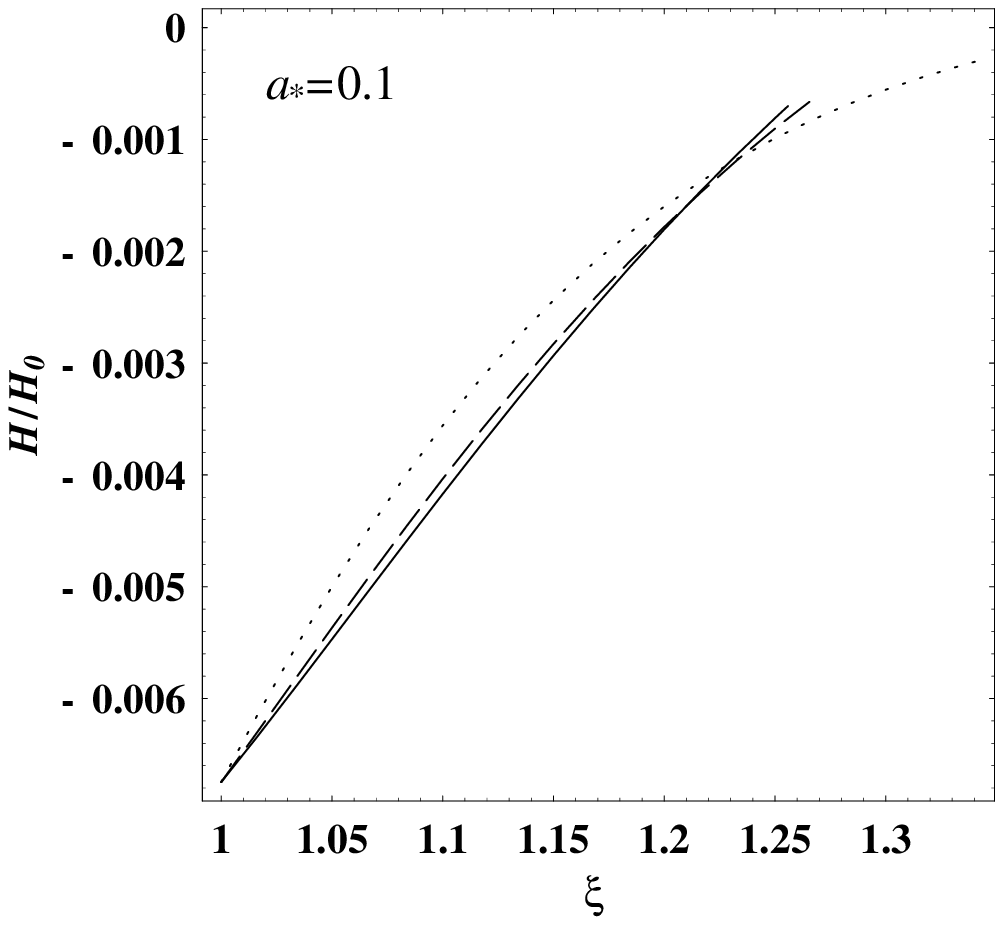}
 \hfill
 \includegraphics[width=5.6cm]{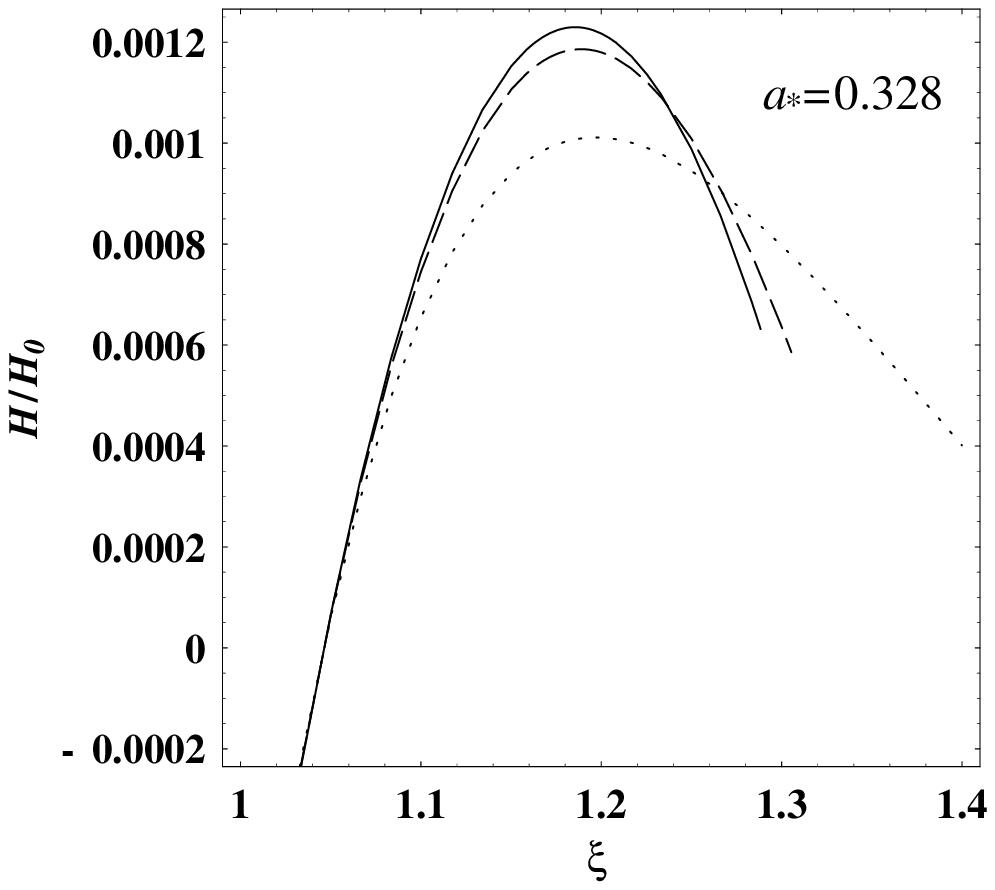}
 \hfill
 \includegraphics[width=5.1cm]{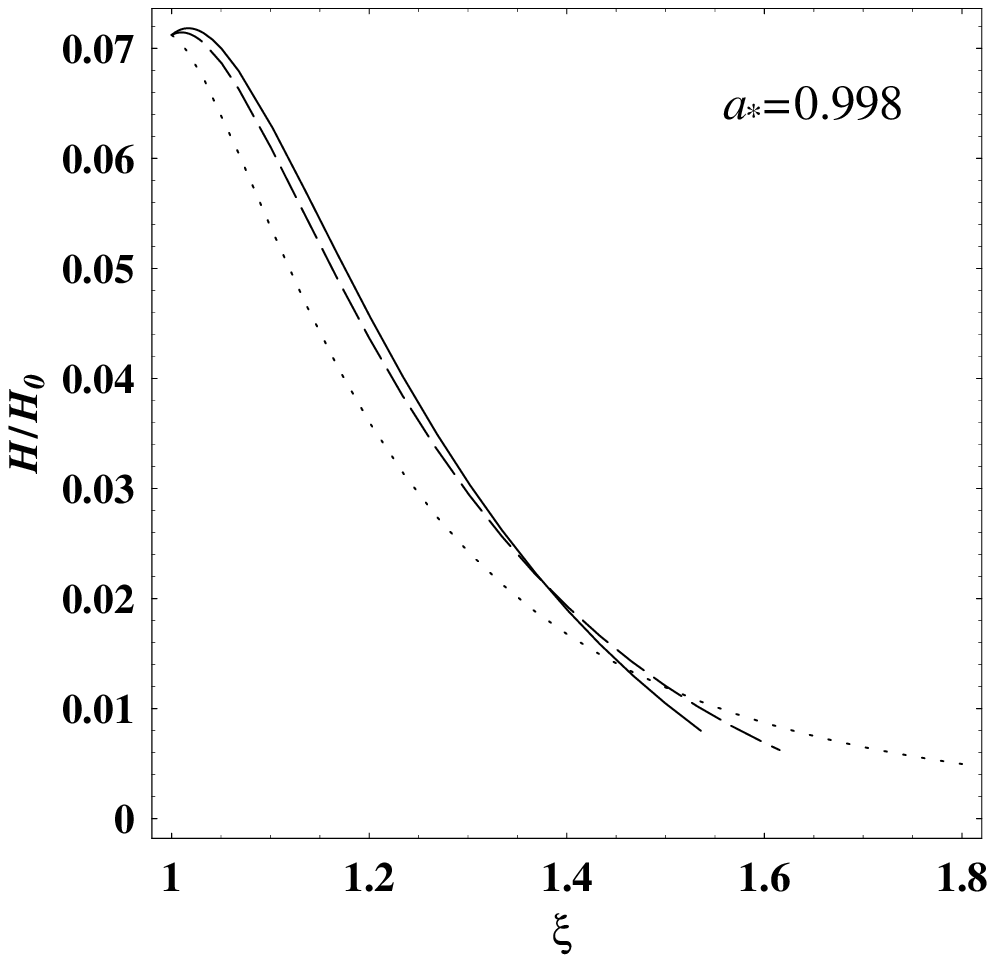}}
 \centerline{\hspace{1.05cm}(a)\hspace{5.9cm}(b)\hspace{5.6cm}(c)}
 \caption{ The curves of ${H(a_ * ;\xi ,n)/H_0}$  versus
$\xi $ for $1 < \xi < \xi _{out} $ with $n = 1.1, 1.5$ and $3.0$
in solid, dashed and dotted lines, respectively. (a)$a_* = 0.1$,
(b) $a_ * = 0.328$, (c) $a_ * = 0.998$.}
 \label{fig6}
 \end{center}
 \end{figure*}

As the magnetic field on the BH is supported by the surrounding
disc, there are some relations between $B_H $ and $\dot {M}_D $.
As a matter of fact these relations might be rather complicated,
and would be very different in different situations. One of them
is given to investigate the correlation between BH spin and
dichotomy of quasars by considering the balance between the
pressure of the magnetic field on the horizon and the ram pressure
of the innermost parts of an accretion flow (Moderski, Sikora {\&}
Lasota 1997), i.e.,

\begin{equation}
\label{eq40} {B_H^2 } \mathord{\left/ {\vphantom {{B_H^2 } {\left(
{8\pi } \right)}}} \right. \kern-\nulldelimiterspace} {\left(
{8\pi } \right)} = P_{ram} \sim \rho c^2\sim {\dot {M}_D }
\mathord{\left/ {\vphantom {{\dot {M}_D } {\left( {4\pi r_{_H}^2 }
\right)}}} \right. \kern-\nulldelimiterspace} {\left( {4\pi
r_{_H}^2 } \right)}.
\end{equation}
From equation (\ref{eq40}) we assume the relation as

\begin{equation}
\label{eq41} \dot {M}_D = {B_H^2 M^2\left( {1 + q} \right)^2}
\mathord{\left/ {\vphantom {{B_H^2 M^2\left( {1 + q} \right)^2}
2}} \right. \kern-\nulldelimiterspace} 2.
\end{equation}
Substituting equations (\ref{eq4}), (\ref{eq5}) and (\ref{eq41})
into equations (\ref{eq1})---(\ref{eq3}), we have

\begin{equation}
\label{eq42} {dM} \mathord{\left/ {\vphantom {{dM} {dt}}} \right.
\kern-\nulldelimiterspace} {dt} = f_{_{MCDA}} \left( {a_ *
,n,\theta _1 } \right)\dot {M}_D,
\end{equation}

\begin{equation}
\label{eq43} {dJ} \mathord{\left/ {\vphantom {{dJ} {dt}}} \right.
\kern-\nulldelimiterspace} {dt} = Mh_{_{MCDA}} \left( {a_ *
,n,\theta _1 } \right)\dot {M}_D,
\end{equation}

\begin{equation}
\label{eq44} {da_ * } \mathord{\left/ {\vphantom {{da_ * } {dt}}}
\right. \kern-\nulldelimiterspace} {dt} = M^{ - 1}g_{_{MCDA}}
\left( {a_ * ,n,\theta _1 } \right)\dot {M}_D,
\end{equation}

\noindent where $f_{_{MCDA}} \left( {a_ * ,n,\theta _1 } \right)$,
$h_{_{MCDA}} \left( {a_ * ,n,\theta _1 } \right)$ and $g_{_{MCDA}}
\left( {a_ * ,n,\theta _1 } \right)$ are the characteristics
functions of BH evolution in the MCDA.

\subsection{Efficiency of the BH-disc system}

The total efficiency of converting accreted mass into radiation
energy corresponding to the BH evolution in MCDA is

\begin{equation}
\label{eq45} \eta _s = 1 - {\left( {{dM} \mathord{\left/
{\vphantom {{dM} {dt}}} \right. \kern-\nulldelimiterspace} {dt}}
\right)} \mathord{\left/ {\vphantom {{\left( {{dM} \mathord{\left/
{\vphantom {{dM} {dt}}} \right. \kern-\nulldelimiterspace} {dt}}
\right)} {\dot {M}_D }}} \right. \kern-\nulldelimiterspace} {\dot
{M}_D } = 1 - f_{_{MCDA}} \left( {a_ * ,n,\theta _1 } \right),
\end{equation}

\noindent and it consists of two parts due to the MC process and
disc accretion as follows:

\begin{equation}
\label{eq46} \eta _s = \eta _{_{DA}} + \eta _{_{MC}}
\end{equation}

\noindent where

\begin{equation}
\label{eq47} \eta _{_{DA}} = 1 - E_{ms} , \quad {\eta _{_{MC}} =
P_{MC} } \mathord{\left/ {\vphantom {{\eta _{_{MC}} = P_{MC} }
{\dot {M}_D }}} \right. \kern-\nulldelimiterspace} {\dot {M}_D }.
\end{equation}
Substituting equations (\ref{eq4}) and (\ref{eq41}) into equation
(\ref{eq47}), we express $\eta _{_{MC}} $ as

\begin{equation}
\label{eq48} \eta _{_{MC}} \left( {a_ * ,n} \right) =
\frac{4\left( {1 - q} \right)}{\left( {1 + q} \right)}\int_1^{\xi
_{out} } {\frac{\beta \left( {1 - \beta } \right)G\left( {a_ *
;\xi ,n} \right)}{2\csc ^2\theta - \left( {1 - q} \right)}} d\xi.
\end{equation}

\noindent It is noticed that $\eta _{_{MC}} $ only depends on the
parameters $a_ * $ and $n$ in our model. The curves of $\eta
_{_{DA}} $ and $\eta _{_{MC}} $ versus $a_ * $ for the given $n$
and $\theta _1 $ are shown in Fig.5.

 From Fig.5 we find the following results for the
efficiency in the MCDA:

(i)The efficiency $\eta _{_{DA}} $ increases monotonically with
$a_* $, while $\eta _{_{MC}} $ varies non-monotonically with $a_ *
$, attaining a maximum for a high BH spin.

(ii)The efficiency $\eta _{_{MC}} $ is greater than $\eta _{_{DA}}
$ for some value range of the high BH spin as listed in Table 5.

(iii)$\eta _{_{MC}} = 0$ occurs just at $a_ * ^P $ for $P_{MC} =
0$, and $\eta _{_{MC}} < 0$ occurs for $0 < a_ * < a_ * ^P $,
implying that energy is transferred from the disc to the BH by the
magnetic field.

(iV)The efficiency $\eta _{_{MC}} $ decreases very rapidly from
its maximum to zero after $a_ * $ passes across the maximum point.

(v) Although $\eta _{_{MC}} $ is not sensitive to the variation of
the parameter $n$, the values of $\left( {\eta _{_{MC}} }
\right)_{\max } $ and the corresponding $\left( {a_ * ^\eta }
\right)_{\max } $ become greater as $n$ increases.

 Some specific value ranges of $\eta _{_{MC}} $ and the
corresponding $a_ * $ are listed in Table 5.

\subsection{ MC effects on disc radiation and transfer of angular
momentum}

Based on the three conservation laws of mass, energy and angular
momentum the following equation of radiation from a thin disc was
derived by considering the MC effects in Li02a:

\begin{equation}
\label{eq49} F = F_{DA} + F_{MC},
\end{equation}

\noindent where $F_{DA} $ is the radiation flux due to disc
accretion as given by Page and Thorne (1974, hereafter PT74):

\begin{equation}
\label{eq50} F_{DA} = {\dot {M}_D f_{DA} } \mathord{\left/
{\vphantom {{\dot {M}_D f_{DA} } {\left( {4\pi r} \right)}}}
\right. \kern-\nulldelimiterspace} {\left( {4\pi r} \right)},
\end{equation}

\noindent and $F_{MC} $ is the radiation flux due to the MC
effects and expressed by

\begin{equation}
\label{eq51}
\begin{array}{l}
F_{MC} = - \frac{d\Omega _D }{rdr}\left( {E^ + - \Omega _D L^ + }
\right)^{ - 2} \times \\ \\
\quad\quad\quad\quad \int_{r_{ms} }^r {\left( {E^+ - \Omega _D L^+
} \right)Hrdr} .
\end{array}
\end{equation}

\noindent Function $H$ in equation (\ref{eq51}) is the flux of
angular momentum transferred between the BH and the disc by the
magnetic field. $E^+$ and $L^+$ are the specific energy and
angular momentum of a particle in the disc, respectively, and read
$\quad $( Novikov {\&} Thorne 1973)

\begin{equation}
E^+=\frac{1-2\chi^{-2}+a_*\chi^{-3}}{(1-3\chi^{-2}+2a_*\chi^{-3})^{1/2}},
\end{equation}

\begin{equation}
L^+=\frac{M\chi(1-2a_*\chi^{-3}+a_*^2\chi^{-4})}{(1-3\chi^{-2}+2a_*\chi^{-3})^{1/2}}.
\end{equation}

\noindent and we have $E^+=E_{ms}$ and $L^+=L_{ms}$ for $\xi=1$
with $\chi=\chi_{ms}$.

 Very recently Li pointed out that the magnetic coupling
between a black hole and a disc can produce a very steep
emissivity with index $\alpha = 4.3 \sim 5.0$, which is consistent
with the recent \textit{XMM-Newton} observation of the nearby
bright Seyfert 1 galaxy MCG-6-30-15 (Li02b). However such a steep
emissivity is very difficult to be explained by a standard
accretion. The emissivity index is defined as

\begin{equation}
\label{eq52} \alpha \equiv - {d\ln F} \mathord{\left/ {\vphantom
{{d\ln F} {d\ln r}}} \right. \kern-\nulldelimiterspace} {d\ln r},
\end{equation}

\noindent which mimics $F \propto r^{ - \alpha }$ locally. In
Li02b the calculation for the emissivity index is done for a
stable non-accretion disc, and the flux function $H$ is assumed to
be distributed from $r = r_{ms} $ to $r = r_{out} $ with a power
law:

\begin{equation}
\label{eq53} H = \left\{ {\begin{array}{l}
 Ar^m,\quad\quad \mbox{ }r_{ms} < r < r_{out} \\ \\
 0,\quad\quad\quad\quad \mbox{ }r > r_{out} \\
 \end{array}} \right.
\end{equation}

\noindent where $A$ is regarded as a constant in Li02b.

Equation (\ref{eq53}) can be modified by using the mapping
relation in our model. From the conservation of angular momentum
and equation (\ref{eq5}) we have

\begin{equation}
\label{eq54} {\partial T_{MC} } \mathord{\left/ {\vphantom
{{\partial T_{MC} } {\partial r}}} \right.
\kern-\nulldelimiterspace} {\partial r} = 4\pi rH = - \frac{4T_0
a_ * \left( {1 + q} \right)\left( {1 - \beta } \right)\sin
^3\theta }{2 - \left( {1 - q} \right)\sin ^2\theta }\frac{\partial
\theta }{\partial r},
\end{equation}

\begin{figure}
\begin{center}
{\includegraphics[width=5.5cm]{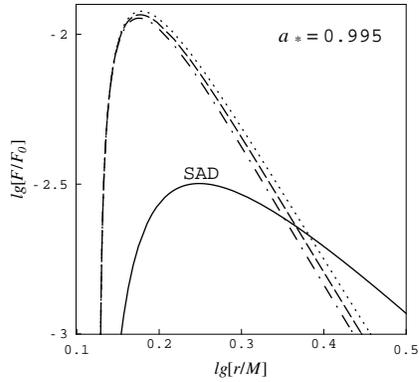}
 \centerline{\hspace{0.8cm}(a)}
 \includegraphics[width=5.5cm]{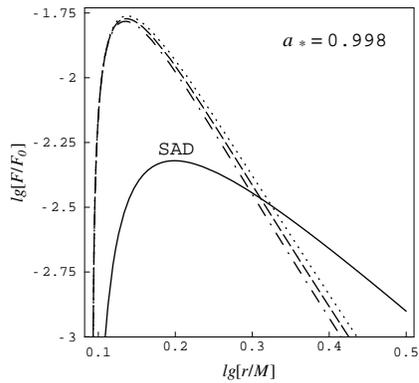}
 \centerline{\hspace{1cm}(b)}}
 \caption{The curves of $\lg(
F_{MC}/F_0)$ versus $\lg(r/M)$ with $\xi _{out} = 2.5$ and $n =
6.5, 7$ and $7.5$ in dotted , dashed and dotted-dashed lines,
respectively. The curve of $\lg (F_{DA}/F_0)$ versus $\lg(r/M)$ is
plotted by a solid line indicating a standard accretion disc
(SAD): (upper panel) $a_* = 0.995$, (lower panel) $a_* = 0.998$.}
\label{fig7}
\end{center}
\end{figure}

\noindent where

\begin{equation}
\label{eq55} {\partial \theta } \mathord{\left/ {\vphantom
{{\partial \theta } {\partial r}}} \right.
\kern-\nulldelimiterspace} {\partial r} = \left( {{\partial \theta
} \mathord{\left/ {\vphantom {{\partial \theta } {\partial \xi }}}
\right. \kern-\nulldelimiterspace} {\partial \xi }} \right)\left(
{{\partial \xi } \mathord{\left/ {\vphantom {{\partial \xi }
{\partial r}}} \right. \kern-\nulldelimiterspace} {\partial r}}
\right) = - \frac{\mbox{G}\left( {a_ * ;\xi ,n} \right)}{r_{ms}
\sin \theta }.
\end{equation}
Substituting equations (\ref{eq18}) and (\ref{eq55}) into equation
(\ref{eq54}), we have

\begin{equation}
\label{eq56}
 H(a_ *;\xi,n) /H_0  =
  \left\{ {\begin{array}{c}
  A(a_ * ,\xi)\xi ^{-n},\quad\quad 1 < \xi <\xi_{out} \\ \\
 0\quad, \quad\quad\quad\quad \xi > \xi _{out} \\
 \end{array}} \right.
\end{equation}

\noindent where we have $H_0 = \left\langle {B_H^2 } \right\rangle
M = 1.48\times 10^{21}\times B_4^2 M_8 \mbox{ }g \cdot s^{ - 2}$,
and

\begin{equation}
\label{eq57}
\left\{ {\begin{array}{l}
 A(a_ * ,\xi) = \frac{a_ * ( 1 - \beta )(
{1 + q} )}{2\pi \chi _{ms}^2 \left[ {2\csc ^2\theta - ( 1 - q)}
\right]}F_A ( a_ * ,\xi ); \\ \\
 F_A (a_ * ,\xi ) =\\
\quad\quad\quad\quad \frac{\sqrt {1 + a_ * ^2 \chi _{ms}^{ - 4}
\xi ^{ - 2} + 2a_
* ^2 \chi _{ms}^{ - 6} \xi ^{ - 3}} }{\sqrt {( 1 + a_ * ^2 \chi
_{ms}^{ - 4} + 2a_ * ^2 \chi _{ms}^{ - 6} )( 1 - 2\chi _{ms}^{ -
2} \xi ^{ - 1} + a_ * ^2 \chi _{ms}^{ - 4} \xi
^{ - 2})} }. \\
 \end{array}} \right.
\end{equation}
Since the magnetic field decreases with increasing $r$, the
power-law index in equation (\ref{eq53}) should be negative,
obeying $m = - n$. Thus we derive a modified expression
(\ref{eq56}) for the function $H$, where the coefficient $A\left(
{a_ * ,\xi } \right)$ is dependent on $a_ * $ and $\xi $ rather
than a constant. For the given values of $a_ * $ and $n$ we have
the curves of ${H\left( {a_ * ;\xi ,n} \right)} \mathord{\left/
{\vphantom {{H\left( {a_ * ;\xi ,n} \right)} {H_0 }}} \right.
\kern-\nulldelimiterspace} {H_0 }$ versus the radial coordinate
$\xi $ for $1 < \xi < \xi _{out} $ as shown in Fig.6.


\begin{table*}
\centerline{\textbf{Table 6.}\it{ The MC parameters adapting the
emissivity index to the observations}}
\begin{tabular}{cccccc}
\hline $\theta _1 $& $a_ * = 0.990$& $a_ * = 0.992$& $a_ * =
0.994$& $a_ * = 0.996$&
$a_ * = 0.998$ \\
\hline $\pi \mathord{\left/ {\vphantom {\pi {12}}} \right.
\kern-\nulldelimiterspace} {12}$& 2.340& 2.329& 2.326& 2.343&
2.423 \\
\hline $\pi \mathord{\left/ {\vphantom {\pi 4}} \right.
\kern-\nulldelimiterspace} 4$& 3.335& 3.339& 3.359& 3.413&
3.576 \\
\hline ${2\pi } \mathord{\left/ {\vphantom {{2\pi } 5}} \right.
\kern-\nulldelimiterspace} 5$& 7.563& 7.673& 7.851& 8.176&
8.944 \\
\hline
\end{tabular}
\label{tab6}
\end{table*}

Inspecting Fig.6, we have the following results on the transfer of
angular momentum:

(i) The value of ${H\left( {a_ * ;\xi ,n} \right)} \mathord{\left/
{\vphantom {{H\left( {a_ * ;\xi ,n} \right)} {H_0 }}} \right.
\kern-\nulldelimiterspace} {H_0 }$ is negative for very low BH
spin as shown in Fig.6(a), and its value is positive for a very
high BH spin  as shown in Fig.6(c).

(ii) As shown in Fig.6(b), positive ${H\left( {a_ * ;\xi ,n}
\right)} \mathord{\left/ {\vphantom {{H\left( {a_ * ;\xi ,n}
\right)} {H_0 }}} \right. \kern-\nulldelimiterspace} {H_0 }$
coexists with negative $H(a_*;\xi,n)/H_0$ for BH spin, such as $a_
* = 0.328$, which is consistent with the value for the coexistence
of the OMCR and the IMCR as shown in Table 3.

(iii) As shown in Fig.6b and 6c, the flux of angular momentum
transferred decreases as $\xi $ approaches $\xi _{out} $.

By using equations (\ref{eq50}), (\ref{eq51}) and (\ref{eq56}) we
have the curves of $\lg(F_{MC}/F_0)$ and $\lg(F_{DA}/F_0)$ versus
$\lg(r/M)$ as shown in Fig.7, where $F_0 $ is defined as $F_0
\equiv \dot {M}_D / r_{ms}^2 $.

Combining equations (\ref{eq51}), (\ref{eq52}) and (\ref{eq56}),
we have the curves of the emissivity index $\alpha $ versus $\lg
\left( {r \mathord{\left/ {\vphantom {r M}} \right.
\kern-\nulldelimiterspace} M} \right)$ as shown in Fig.8.

From Fig.7 we find that $\lg \left( {{F_{MC} } \mathord{\left/
{\vphantom {{F_{MC} } {F_0 }}} \right. \kern-\nulldelimiterspace}
{F_0 }} \right)$ varies much more steeply with $\lg \left( {r
\mathord{\left/ {\vphantom {r M}} \right.
\kern-\nulldelimiterspace} M} \right)$ than $\lg \left( {{F_{DA} }
\mathord{\left/ {\vphantom {{F_{DA} } {F_0 }}} \right.
\kern-\nulldelimiterspace} {F_0 }} \right)$ does. It is found in
Fig.8 that we can adapt the emissivity index arising from $ -
{d\ln F_{MC} } \mathord{\left/ {\vphantom {{d\ln F_{MC} } {d\ln
r}}} \right. \kern-\nulldelimiterspace} {d\ln r}$ to the
observations of the observations by the curves in the shaded
region, while the index of a standard accretion disc (SAD) is far
below the shaded region. This result is one of the observational
signatures of the magnetic coupling between a rotating BH and its
surrounding disc. Compared with the model given in Li02b we have
more parameters to choose for adapting the emissivity index to the
observations. Considering both the mapping relation (\ref{eq19})
and the observations of the Seyfert 1 galaxy MCG-6-30-15, we have
the values of the power-law index $n$ corresponding to the
different values of $\theta _1 $ and $a_ * $ as shown in Table 6.

From Table 6 we find that the values of $n$ adapted to the
observations are generally less than those given in Li02b, if
$\theta _1 $ is not close to $\pi \mathord{\left/ {\vphantom {\pi
2}} \right. \kern-\nulldelimiterspace} 2$. This result implies
that the concentration of the magnetic field on the disc could be
relaxed in our model.

It has been proved in PT74 that the internal viscous torque per
unit circumference $W_\varphi ^r $ is related to the radiation
flux $F$ by

\begin{equation}
\label{eq58} W_\varphi ^r = \frac{2\left( {E^ + - \Omega _D L^ + }
\right)}{\left( {{ - d\Omega _D } \mathord{\left/ {\vphantom {{ -
d\Omega _D } {dr}}} \right. \kern-\nulldelimiterspace} {dr}}
\right)}F.
\end{equation}

\begin{figure}
\begin{center}
{\includegraphics[width=5.5cm]{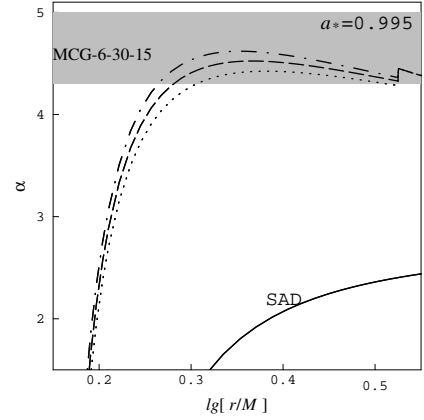}
 \centerline{\hspace{0.6cm}(a)}
 \includegraphics[width=5.5cm]{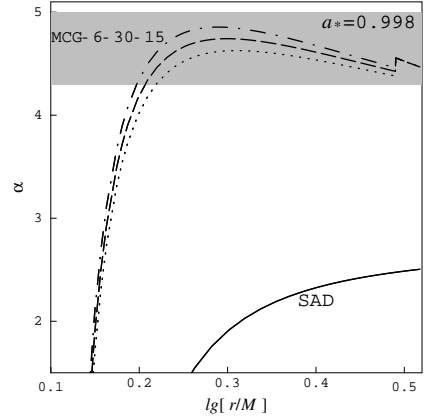}
\centerline{\hspace{0.6cm}(b)}}
 \vspace{0.1cm}
 \caption{ The emissivity index $\alpha$ versus $\lg ( r/M)$ with $\xi _{out} =
2.5$ and $n = 6,\mbox{ }7$ and $8$ in dotted, dashed and
dot-dashed lines, respectively. The emissivity index of the
Seyfert 1 galaxy MCG-6-30-15 inferred from the observation of
{\textit{XMM-Newton}} is shown by the shaded region. The
emissivity index of a standard accretion disc (SAD) is plotted in
solid line. (upper panel) $a_ * = 0.995$ , (lower panel) $a_ * =
0.998$.}
 \label{fig8}
 \end{center}
\end{figure}

\noindent Equation (\ref{eq58}) was derived based on the three
laws of conservation, and it is proved to be valid for the MC
process in Li02a. Combining equations (\ref{eq49}) and
(\ref{eq58}), we can express the contribution to $W_\varphi ^r $
as follows:

\begin{equation}
\label{eq59}
W_\varphi ^r = \left( {W_\varphi ^r } \right)_{DA} +
\left( {W_\varphi ^r } \right)_{MC} ,
\end{equation}

\begin{figure*}
\begin{center}
{\includegraphics[width=5.3cm]{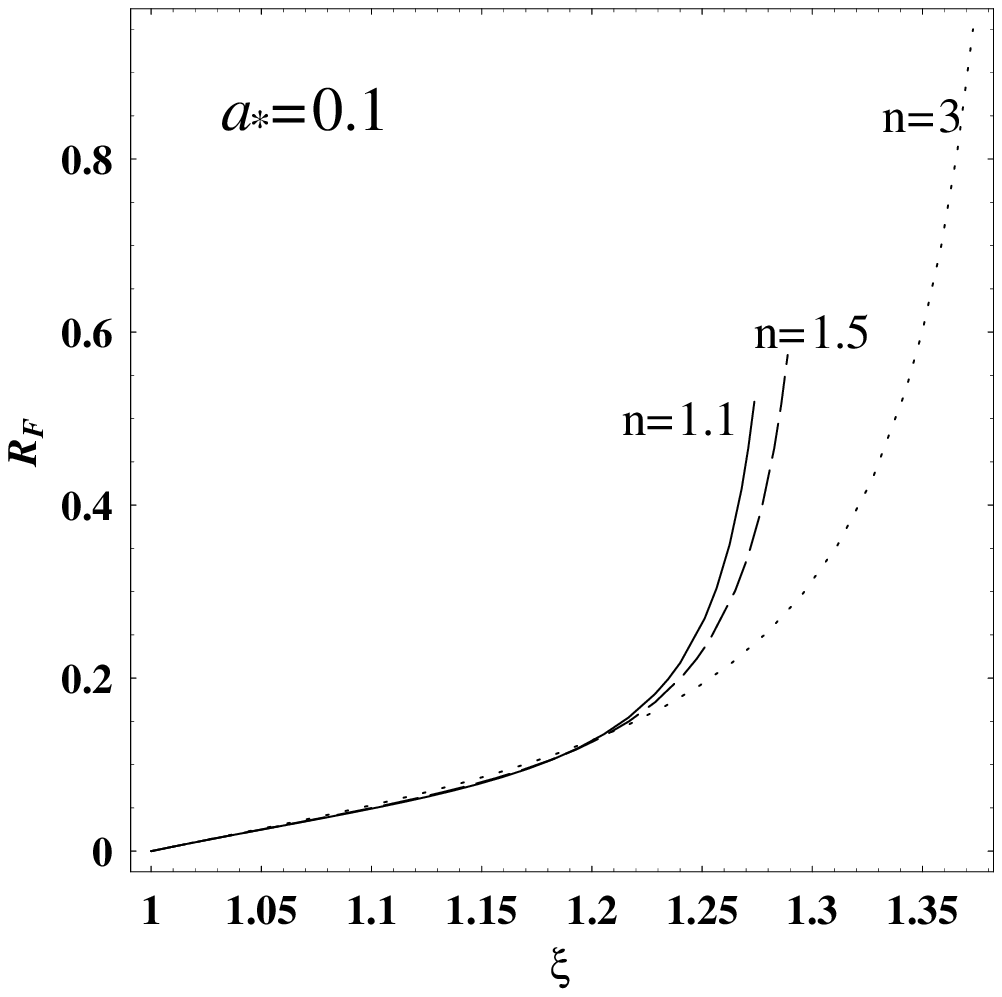}
 \hfill
 \includegraphics[width=5.5cm]{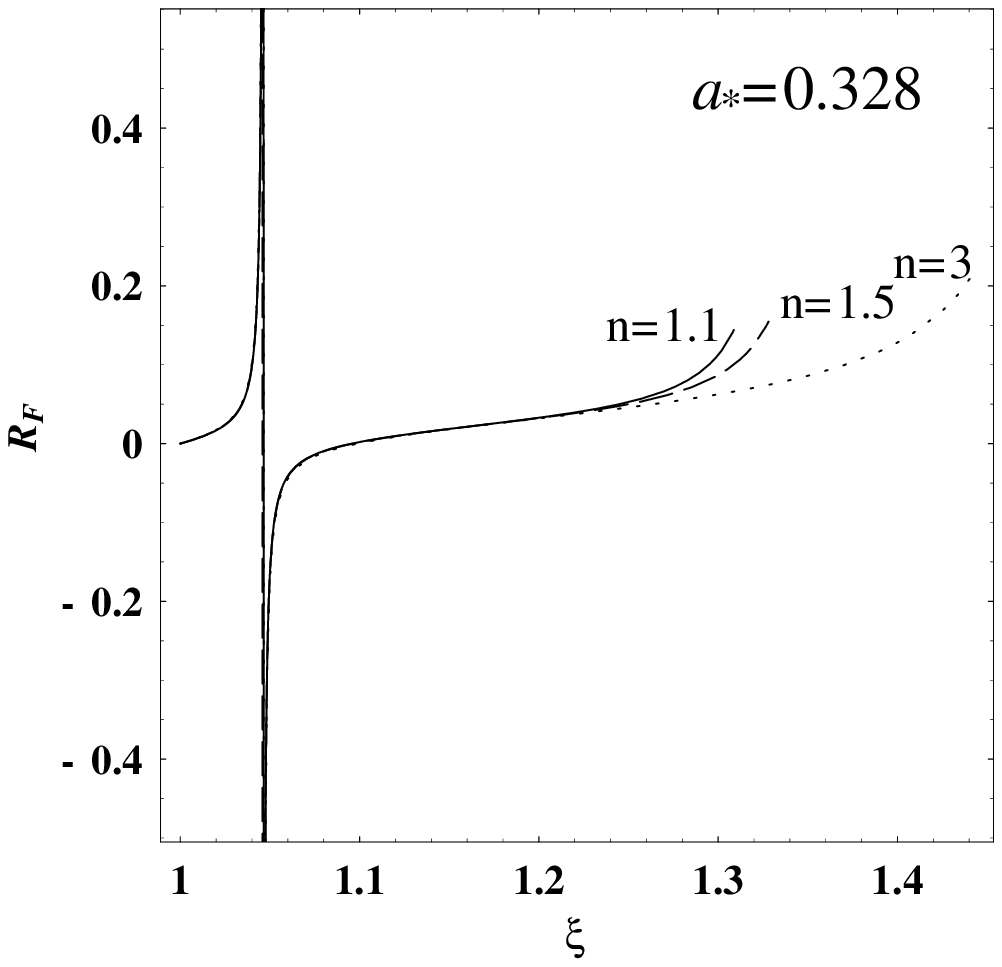}
 \hfill
 \includegraphics[width=5.4cm]{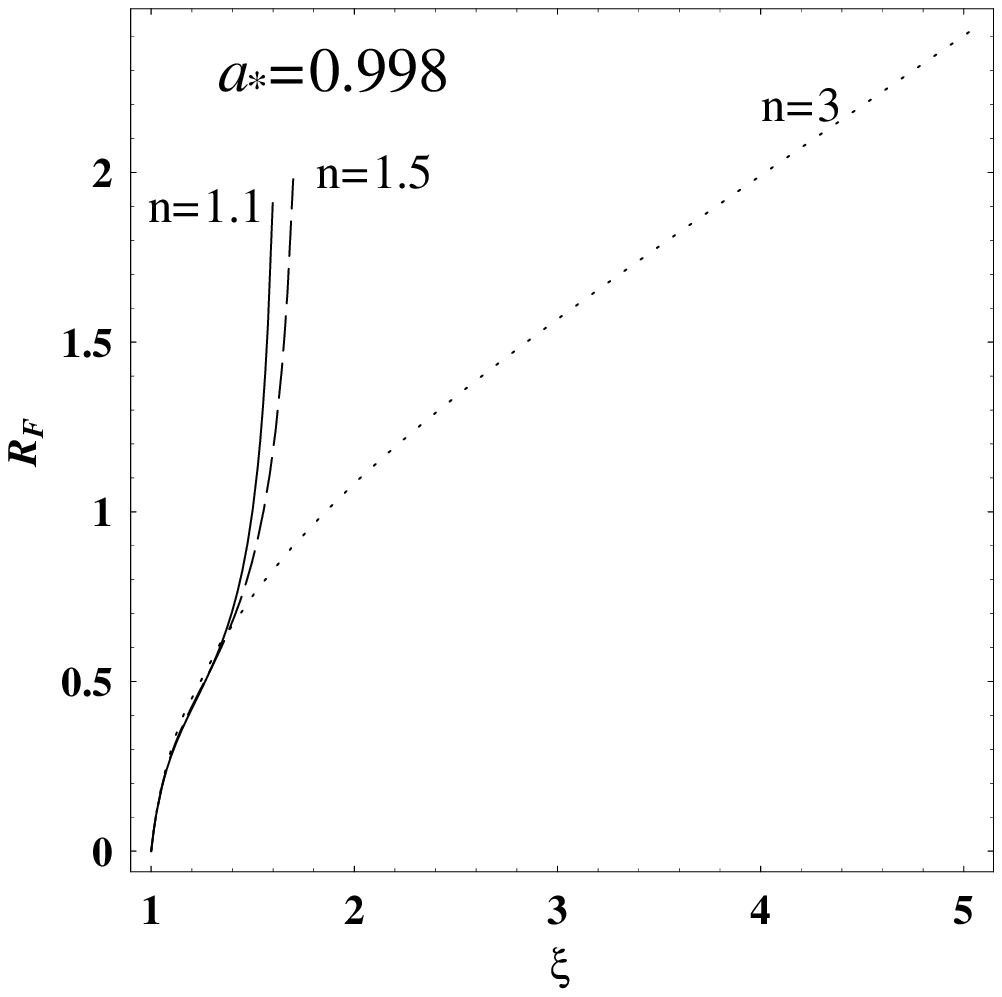}}
\centerline{\hspace{0.8cm}(a)\hspace{5.8cm} (b)\hspace{5.7cm} (c)}
 \caption{ The curves of $R_F $ versus
$\xi $ for $1 < \xi < \xi _{out} $ with $n = 1.1, 1.5$ and $3.0$
in solid, dashed and dotted lines, respectively. (a) $a_ * = 0.1$,
(b) $a_ * = 0.328$, (c) $a_ * = 0.998$.}
 \label{fig9}
\end{center}
\end{figure*}

\noindent where $\left( {W_\varphi ^r } \right)_{DA} $ and $\left(
{W_\varphi ^r } \right)_{MC} $ are the contribution due to disc
accretion and the MC process, and are related to $F_{DA} $ and
$F_{MC} $ by equation (\ref{eq58}), respectively. Since the flux
of angular momentum transferred away from the disc by radiation
(henceforth FAMFD) is $F_{MC} L^ + $, and that transferred from
the BH into the disc by the magnetic field (henceforth FAMFH) is
$H$, the ratio of the two can be expressed as

\begin{equation}
\label{eq60}
 \begin{array}{l}
 R_F \equiv \frac{F_{MC} L^ + }{H} \\ \\
 = - \frac{L^+}{Hr}\frac{d\Omega _D }{dr}\left( {E^ + - \Omega _D L^ + }
\right)^{ - 2}\int_{r_{ms} }^r {\left( {E^ + - \Omega _D L^ + }
\right)Hrdr}.
\end{array}
\end{equation}
From equations (\ref{eq60}) we obtain the curves of $R_F $ versus
$\xi $ for the given values of $n$ and $a_ * $ as shown in Fig.9.

Combining Figs.6 and 9, we obtain the following relations between
FAMFD and FAMFH:

(i)As shown in Fig.9(a), we have $0 < R_F < 1$, which means that
the absolute value of FAMFD is less than that of FAMFH, i.e. $H <
F_{MC} L^ + < 0$, for a BH with low spin such as $a_ * = 0.1$.

(ii)As shown in Fig.9(b), we have $0 < R_F < + \infty $ and $ -
\infty < R_F < 0$, and $R_F $ approaches infinity at the place
where $H$ changes its sign. So we infer that $F_{MC} L^ + < 0$
holds in the whole MC region $1 < \xi < \xi _{out} $ by
considering the sign of $H$ in Fig.6(b).

(iii)As shown in Fig.9(c), we have $0 < R_F < 1$ for $\xi _{in} <
\xi < \xi _{eq} $, where FAMFD is dominated by FAMFH. And we have
$R_F
> 1$ for $\xi _{eq} < \xi < \xi _{out} $, where FAMFD dominates
over FAMFH. It is easy to obtain that the radial coordinate $\xi
_{eq} $ indicating $R_F = 1$ is equal to 1.498, 1.552 and 1.857
for $n = $1.1, 1.5 and 3.0, respectively.

From the above discussion we find that both FAMFD and FAMFH are
related intimately not only to the BH spin but also to the disc
location.

\section{SUMMARY}

In this paper the transfer of energy and angular momentum between
a rotating BH and its surrounding accretion disc is discussed in
detail by considering MC effects. Our discussion is given for two
cases: (i)the MC process without disc accretion, (ii)the MC
process with disc accretion. In the first case only the two
conservation laws of energy and angular momentum are used, while
in the second case the conservation of accreted mass is used in
addition. Compared with Li02a and Li02b, the mapping relation
(\ref{eq19}) is used to depict the MC process throughout this
paper, and the correlation of some parameters with MC effects is
discussed, where the BH spin $a_ * $, the power-law index $n$ and
the radial coordinate $\xi $ are involved.

Compared with the BZ process we find that the MC effects do not
increase monotonically as the BH spin. For example, in Figs. 4 and
5 both $\left( {rate} \right)_S $ and $\eta _{_{MC}} $ attain
their maxima and then decrease very rapidly as $a_ * $ approaches
unity. These results can be explained by the equations
(\ref{eq39}) and (\ref{eq48}) with the behavior of $\xi _{out} $
as $a_ * $ approaches unity. From Fig.2 we find that the outer
radius of MC region approaches the inner edge of the disc very
closely as $a_ * $ approaches unity, and accordingly the ratio
$\beta \equiv {\Omega _D } \mathord{\left/ {\vphantom {{\Omega _D
} {\Omega _H }}} \right. \kern-\nulldelimiterspace} {\Omega _H }$
approaches unity along as $r_{ms} $ approaches the horizon
radius $r_{_H} $.\\

\noindent\textbf{Acknowledgments. }This work was supported by the
National Natural Science Foundation of China under Grant No.
10173004 and No. 10121503. The anonymous referee is thanked for
his suggestion on modification of the mapping relation (19).

\end{document}